\documentclass[conference,compsoc]{IEEEtran}

\usepackage{amsthm}
\usepackage{amsmath}
\usepackage{amsfonts}
\usepackage[bb=boondox]{mathalfa}
\usepackage{graphicx}
\usepackage{multirow}
\usepackage{booktabs}
\usepackage{color}
\usepackage{array}
\usepackage{colortbl}
\usepackage{mathrsfs}
\usepackage{enumitem}
\usepackage{amssymb}
\usepackage{pifont}
\usepackage{mathabx}

\usepackage{makecell}
\usepackage{float}
\usepackage{subfigure}
\usepackage{threeparttable}
\usepackage{caption}
\usepackage[linesnumbered,ruled,vlined]{algorithm2e}
\usepackage{graphicx}
\usepackage{nicefrac}
\usepackage{hyperref}

\newcommand{\bs}[1]{\mathbf{#1}}
\newcommand{\opn}[1]{\operatorname{#1}}
\DeclareMathAlphabet{\mymathbb}{U}{BOONDOX-ds}{m}{n}
\newcommand{\cmark}{\ding{51}}
\newcommand{\xmark}{\ding{55}}
\newcommand{\omark}{$\ovoid$}

\newcommand{\rseq}[3]{
    \resizebox{#1}{#2}{
        \begin{math}
            \begin{aligned}
                #3
            \end{aligned}
        \end{math}
    }
}

\newtheorem{assumption}{Assumption}
\newtheorem{lemma}{Lemma}
\newtheorem{theorem}{Theorem}

\ifCLASSOPTIONcompsoc
  \usepackage[nocompress]{cite}
\else
  \usepackage{cite}
\fi

\begin{document}

\date{}

\title{\Large \bf G$^2$uardFL: Safeguarding Federated Learning Against Backdoor Attacks through \\ Attributed Client Graph  Clustering}


\author{\IEEEauthorblockN{Hao Yu$^\dagger$, Chuan Ma$^\ddagger$, Meng Liu$^\dagger$, Tianyu Du$^\star$, Ming Ding$^\S$, Tao Xiang$^\P$, Shouling Ji$^\star$, Xinwang Liu$^\dagger$}
\IEEEauthorblockA{$^\dagger$ \textit{National University of Defense Technology, China} \hspace{0.15in} $^\ddagger$ \textit{Zhejiang Lab, China}}
\IEEEauthorblockA{$^\star$ \textit{Zhejiang University, China} \hspace{0.15in} $^\S$ \textit{Data 61, Australia} \hspace{0.15in} $^\P$ \textit{Chongqing University, China}}
}

\maketitle

\begin{abstract}
Federated Learning (FL) offers collaborative model training without data sharing but is vulnerable to backdoor attacks, where poisoned model weights lead to compromised system integrity.
Existing countermeasures, primarily based on anomaly detection, are prone to erroneous rejections of normal weights while accepting poisoned ones, largely due to shortcomings in quantifying similarities among client models.
Furthermore, other defenses demonstrate effectiveness only when dealing with a limited number of malicious clients, typically fewer than 10\%.
To alleviate these vulnerabilities, we present G$^2$uardFL, a protective framework that reinterprets the identification of malicious clients as an attributed graph clustering problem, thus safeguarding FL systems.
Specifically, this framework employs a client graph clustering approach to identify malicious clients and integrates an adaptive mechanism to amplify the discrepancy between the aggregated model and the poisoned ones, effectively eliminating embedded backdoors.
We also conduct a theoretical analysis of convergence to confirm that G$^2$uardFL does not affect the convergence of FL systems.
Through empirical evaluation, comparing G$^2$uardFL with cutting-edge defenses, such as FLAME (USENIX Security 2022)~\cite{nguyen2022flame} and DeepSight (NDSS 2022)~\cite{rieger2022deep}, against various backdoor attacks including 3DFed (SP 2023)~\cite{li20233dfed}, our results demonstrate its significant effectiveness in mitigating backdoor attacks while having a negligible impact on the aggregated model's performance on benign samples (i.e., the primary task performance).
For instance, in an FL system with 25\% malicious clients, G$^2$uardFL reduces the attack success rate to 10.61\%, while maintaining a primary task performance of 73.05\% on the CIFAR-10 dataset.
This surpasses the performance of the best-performing baseline, which merely achieves a primary task performance of 19.54\%.
\end{abstract}

\section{Introduction}\label{sec:intro}

Federated Learning (FL) has emerged as a novel decentralized learning paradigm that empowers clients to collectively construct a deep learning model (global model) while retaining the confidentiality of their local data.
In contrast to traditional centralized methods, FL requires the transmission of individually trained model weights from clients to a central orchestration server for the aggregation of the global model throughout the process.
FL has gained prominence due to its ability to address critical challenges such as data privacy, resulting in increased adoption across diverse applications, e.g., self-driving cars~\cite{elbir2022federated,pokhrel2020federated}.

However, FL introduces a trade-off in that no client, other than themselves, can access their own data during training.
As prior research has highlighted, this inherent characteristic renders FL susceptible to \emph{backdoor attacks}~\cite{li20233dfed,nguyen2022flame,zhang2022fldetector}.
These attacks involve a minority subset of clients, malicious and compromised by an attacker, transmitting manipulated weights to the server, thereby corrupting the global model.
The objective of these attacks is to induce the global model to make predictions aligned with the attacker's intent for specific samples, often triggered by specific patterns, e.g., car images with racing stripes~\cite{bagdasaryan2020how} or sentences about Athens~\cite{wang2020attack}.

\begin{table}[t]
    \centering
    \caption{Summary of existing defenses.}\label{tab:summary_of_defenses}
    \renewcommand\arraystretch{1.3}
    \scalebox{0.64}{
    \begin{threeparttable}
        \begin{tabular}{m{2.0cm} m{1.2cm}<{\centering} m{1.8cm}<{\centering} m{1.8cm}<{\centering} m{1.8cm}<{\centering} m{1.8cm}<{\centering}}
            \toprule[1.2pt]
            \multicolumn{1}{c}{\textbf{Methods}} & \textbf{Client filtering} & \textbf{Historical information} & \makecell[c]{\textbf{Server's} \\ \textbf{data absence}$^1$} & \textbf{Noise-free aggregation} & \textbf{Poison eliminating} \\
            \midrule[1pt]
            FLAME~\cite{nguyen2022flame} & \cmark & \xmark & \cmark & \xmark & \xmark  \\
            DeepSight~\cite{rieger2022deep} & \cmark & \xmark & \omark$^2$ & \cmark & \xmark \\
            FoolsGold~\cite{fung2020limitations} & \cmark & \cmark & \cmark & \cmark & \xmark \\
            (Multi-)Krum~\cite{blanchard2017machine} & \cmark & \xmark & \cmark & \cmark & \xmark \\
            RLR~\cite{ozdayi2021defending} & \xmark & \xmark & \cmark & \cmark & \xmark \\
            FLTrust~\cite{cao2021fltrust} & \xmark & \xmark & \xmark & \cmark & \xmark \\
            NDC~\cite{sun2019can} & \xmark & \xmark & \cmark & \cmark & \xmark \\
            RFA~\cite{pillutla2019robust} & \cmark & \xmark & \cmark & \xmark & \xmark \\
            Weak DP~\cite{dwork2014algorithmic} & \xmark & \xmark & \cmark & \xmark & \xmark \\
            G$^2$uardFL (ours) & \cmark & \cmark & \cmark & \cmark & \cmark \\
            \bottomrule[1.2pt]
        \end{tabular}
        \begin{tablenotes}
            \item [1]: The orchestration server utilizes its local data to identify malicious clients.
            \item [2]: DeepSight generates random input data to measure changes in model predictions.
        \end{tablenotes}
    \end{threeparttable}
    }
    \vspace{-15pt}
\end{table}

\noindent \textbf{Deficiencies of Existing Defenses}.
Countermeasures against backdoor attacks (Table~\ref{tab:summary_of_defenses}) broadly fall into two categories.
The first category encompasses anomaly detection methods~\cite{nguyen2022flame,rieger2022deep} designed to identify malicious clients by detecting anomalous ones within the distribution of client weights.
However, these solutions are often prone to the erroneous rejection of normal weights or the acceptance of poisoned ones due to inadequate client representations, e.g., cosine similarity~\cite{fung2020limitations,nguyen2022flame} or Euclidean distance~\cite{blanchard2017machine} between model updates.
The second category explores anomaly detection-free methods that employ various techniques, e.g., robust aggregation rules~\cite{ozdayi2021defending,pillutla2019robust}, differential privacy~\cite{dwork2014algorithmic,sun2019can} and local validation~\cite{cao2021fltrust}, to mitigate the influence of poisoned weights during aggregation.
However, these defenses are primarily effective when the number of malicious clients is relatively small (typically less than 10\%~\cite{nguyen2023backdoor}).
Moreover, some methods, like local validation, assume the server collects a small number of clean samples, which contradicts the essence of FL~\cite{zhang2022flip}, and differential privacy may adversely affect the primary task performance of the global model.
Hence, \emph{the effectiveness of defenses hinges on 1) the precise identification of malicious clients and 2) the successful elimination of the influence of previously embedded backdoors}.

\noindent \textbf{Our Key Idea}.
To address these challenges, this paper introduces G$^2$uardFL, a framework designed to safe\underline{g}uard FL systems against backdoor attacks by redefining the detection of malicious clients as an attributed \underline{g}raph clustering problem.
Prior defenses motivated by traditional clustering algorithms like HDBSCAN rely heavily on the discrimination of model weights when calculating similarities among clients.
Nevertheless, model weights, particularly those associated with textual data, are inherently high-dimensional.
Textual models often entail numerous weights related to word embeddings.
Consequently, relying solely on the direct utilization of model weights proves inadequate for accurately identifying malicious clients (Table~\ref{tab:detection_malicious}).
Differing from these defenses, G$^2$uardFL employs low-dimensional representations for capturing client features rather than relying directly on high-dimensional weights.
Additionally, it introduces three methods to model inter-client relationships, inspired by the analysis of poisoned model weights.
By constructing an attributed graph and employing a graph clustering approach that leverages low-dimensional representations and inter-client relationships, G$^2$uardFL effectively segregates clusters with similar characteristics and close associations.
By combining current and historical detection results (i.e., \emph{benign scores}), malicious client clusters are discerned and effectively filtered out, addressing the first challenge.
Additionally, an adaptive mechanism is proposed to reduce the effectiveness of backdoors by eliminating previously embedded ones, effectively addressing the second challenge.

\noindent \textbf{Our Contributions}.
We formally present G$^2$uardFL, an effective and robust defense that introduces the attributed client graph clustering technique to minimize the impact of backdoor attacks while preserving the primary task performance.
The defense scheme consists of four key components: a \emph{graph construction} module that translates inter-client relations into an attributed graph, a \emph{graph clustering} module that assigns clients to distinct clusters, a \emph{dynamic filtering and clipping} module that identifies malicious clients and limits their influence, and an \emph{adaptive poison eliminating} module that removes previously implanted backdoors.
The main contributions can be summarized as follows:
\begin{itemize}[leftmargin=10pt]
    \item To the best of our knowledge, our study pioneers the reformation of defense strategies against backdoor attacks by redefining it as an attributed graph clustering problem. We demonstrate that directly applying the similarity of high-dimensional weights of client models does not fully exploit the model information, thereby resulting in a drop in the accuracy of anomaly detection.
    \item We introduce G$^2$uardFL, a defense framework based on attributed client graph clustering, designed to accurately detect malicious clients under various backdoor attacks and employ an adaptive mechanism to amplify divergence between normal and poisoned models. Additionally, we provide a theoretical convergence analysis, proving that G$^2$uardFL does not affect the convergence of FL systems.
    \item Thorough evaluations of diverse datasets demonstrate that G$^2$uardFL surpasses state-of-the-art solutions across various FL backdoor attack scenarios.
    Specifically, G$^2$uardFL effectively identifies malicious clients and significantly reduces the attack success rate to below 12\% on these datasets, outperforming the best-performing baseline.
\end{itemize}

\section{Background}\label{sec:back}

\subsection{Federated Learning}\label{subsec:federated_learning}

FL is an emerging approach for training decentralized machine learning models, denoted as $G$, across a multitude of clients $\mathcal{C}$ and an orchestration server $\mathcal{S}$~\cite{mcmahan2017communication}.
In each communication round $r \in [1, R]$, the server $\mathcal{S}$ selects a random subset of $m$ clients $\mathcal{C}_m^r$ and shares the current global model $G^r$ with them.
The choice of $m$ aims to balance communication costs per round and overall training time.
Subsequently, each client $c_i \in \mathcal{C}_m^r$ initializes its local model $W_{c_i}^r$ with $G^r$ and fine-tunes it using its local data $\mathcal{D}_{c_i}$.
Following training, each client transmits its updated local model $W_{c_i}^r$ back to the server $\mathcal{S}$, which then employs an aggregation mechanism to construct the new global model $G^{r + 1}$.
Throughout this process, multiple clients collaboratively train a shared model while preserving the privacy of their local data, not sharing it with the server.
Key notations used in this paper are summarized in Table~\ref{tab:notations}.

Various aggregation mechanisms have been proposed, including FedAvg~\cite{mcmahan2017communication} and Krum~\cite{blanchard2017machine}.
This paper primarily focuses on the FedAvg mechanism, as it is widely applied in FL~\cite{elbir2022federated,hard2018federated,pokhrel2020federated} and related works on backdoor attacks~\cite{ma2020safe,ma2022when,nguyen2022flame,sun2019can,wang2020attack}.
In FedAvg, the global model $G^{r + 1}$ at round $r + 1$ is computed as a weighted average of client weights, given by $G^{r + 1} = \sum_{c_i \in \mathcal{C}_m^r} ( \nicefrac{n_{c_i}}{n} ) W_{c_i}^r$, where $n_{c_i} = \lVert \mathcal{D}_{c_i} \rVert$ represents the number of samples, and $n = \sum_{c_i \in \mathcal{C}_m^r} n_{c_i}$.
We also evaluate the performance of Krum under backdoor attacks.

\subsection{Backdoor Attacks on FL}\label{subsec:backdoor_on_federated}

The susceptibility of FL to backdoor attacks has been recognized~\cite{bagdasaryan2020how}, making these attacks a significant security concern.
In a backdoor attack, an attacker $\mathcal{A}$ compromises a subset of $K$ clients and manipulates their weights, changing them from $W_{c_i}$ to $\tilde{W}_{c_i}$ ($c_i \in \mathcal{C}_K$).
These manipulated weights may then be aggregated into the global model $G$, impacting its performance.
The attacker's goal is to induce the global model to make specific, attacker-determined predictions on samples $x_{\mathcal{A}} \sim P_{\mathcal{D}^{\mathcal{A}}}$, where $\mathcal{D}^{\mathcal{A}}$ represents the set of backdoored samples.
Backdoor attacks can be classified into two types based on their execution stages: \emph{data poisoning}~\cite{nguyen2023backdoor,nguyen2020poisoning,wang2020attack,wenger2021backdoor} and \emph{model poisoning}~\cite{bagdasaryan2020how,bhagoji2019analyzing,li20233dfed,sun2019can,zhang2022neurotoxin}.

\noindent \textbf{Data Poisoning}.
In data poisoning attacks, the attacker has full control over the training data collection process for malicious clients, enabling the injection of backdoored samples into the original training data.
These backdoored samples can be further categorized into semantic samples, which exhibit specific properties or are out-of-distribution (e.g., cars with striped patterns~\cite{bagdasaryan2020how} or uncommon airplane types~\cite{wang2020attack}), and artificial samples, which have specific triggers (e.g., added square patterns in corners~\cite{wenger2021backdoor}).
Backdoored samples created through data poisoning serve as a basis for constructing poisoned weights.

\noindent \textbf{Model Poisoning}.
Data poisoning attacks are often hindered by the aggregation mechanism, which mitigates most of the poisoned weights and causes the global model to forget the backdoor~\cite{bagdasaryan2020how,zhang2022neurotoxin}.
To address this, model poisoning attacks directly modify model weights to amplify the influence of poisoned weights.
Model poisoning attacks require the attacker $\mathcal{A}$ to have complete control over the training process.
Bagdasaryan \emph{et al.}~\cite{bagdasaryan2020how} first introduced the model replacement method, which replaces the new global weights $G^{r + 1}$ with poisoned weights by scaling them by a certain factor.
Bhagoji \emph{et al.}~\cite{bhagoji2019analyzing} alternated the objective function to evade the anomaly detection algorithm, preventing the attack from being detected.
Additionally, the projected gradient descent (PGD)~\cite{sun2019can} is employed to project poisoned weights $\tilde{W}_{c_i}$ onto a small sphere centered around global weights $G^r$, ensuring that $\tilde{W}_{c_i}$ does not deviate significantly from $G^r$.
Li \emph{et al.}~\cite{li20233dfed} designed an adaptive and multi-layered framework, 3DFed, to utilize the attack feedback in the preceding epoch for improving the attack performance.

Since data and model poisoning attacks operate at different stages, combining these attack types can enhance the efficacy of backdoor attacks.
In this paper, we assess the effectiveness of G$^2$uardFL under both data and model poisoning attack scenarios.
\section{Threat Model and Design Goals}\label{sec:threat_and_design}

\subsection{Threat Model}\label{subsec:threat}

In the context of countering backdoor attacks, it is crucial to establish clear boundaries for the knowledge and abilities possessed by both attackers and defenders.

\begin{figure*}[t]
    \centering
    \includegraphics[width=0.92\linewidth]{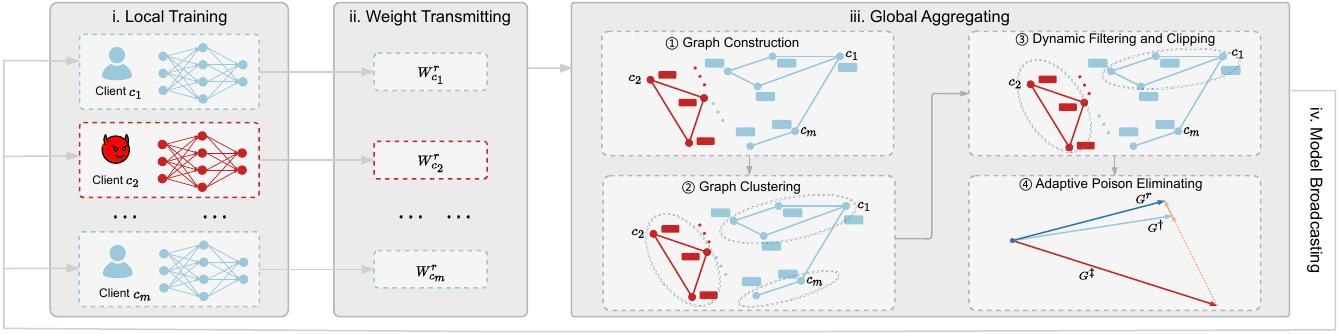}
    \caption{The flowchart illustrates G$^2$uardFL within the FL system.
    Each communication round comprises four sequential steps: local training, weight transmitting, global aggregating, and model broadcasting.}\label{fig:overview}
    \vspace{-14pt}
\end{figure*}

\noindent \textbf{Attacker Knowledge and Abilities}.
Drawing from prior research~\cite{bagdasaryan2020how,wang2020attack,zhang2022flip}, we assume that the attacker $\mathcal{A}$ remains unaware of the orchestration server's inner workings, including the aggregation process and any implemented backdoor defense mechanisms. 
Consequently, the attacker cannot manipulate server operations.
Note that the attacker can devise a specific mechanism to probe inner workings, like the indicators used in 3DFed~\cite{li20233dfed}.

Once the attacker successfully compromises clients, which could be devices such as smartphones, wearable devices, or edge devices~\cite{huang2022cross}, they gain control over all aspects of data collection and model training.
This control extends to hyper-parameters such as the number of training epochs and learning rates.
The attacker is free to execute data poisoning attacks, model poisoning attacks, or a combination of both.
In data poisoning attacks, the attacker modifies backdoored samples to generate poisoned weights.
To maintain a low profile, the attacker employs a Poisoned Data Rate (PDR)~\cite{bagdasaryan2020how,wang2020attack}, which constrains the deviation of poisoned weights from the global model.
Specifically, if $\tilde{\mathcal{D}}_{c_i}$ represents all training samples in a malicious client $c_i$, and $\mathcal{D}_{c_i}^{\mathcal{A}}$ represents the subset of backdoored samples within $\tilde{\mathcal{D}}_{c_i}$, the PDR for client $c_i$ is defined as $PDR_{c_i} = \nicefrac{\lVert \mathcal{D}_{c_i}^{\mathcal{A}} \rVert}{\lVert \tilde{\mathcal{D}}_{c_i} \rVert}$.
We denote the fraction of compromised clients as the Poisoned Model Rate $PMR = \nicefrac{K}{M}$, with the assumption that the attacker can control up to 50\% of clients.
Importantly, due to the server's random selection of a fraction of $M$ clients in each FL round, the exact number of malicious clients participating remains unpredictable.

\noindent \textbf{Defender Knowledge and Abilities}.
Given that clients may include resource-constrained devices like wearable devices, the defender's abilities are constrained to deploying defenses exclusively at the server $\mathcal{S}$.
The server is honest, with no intentions of leaking client information through model weights.
In contrast to FLTrust~\cite{cao2021fltrust}, to safeguard privacy, the server is assumed to possess no local data while possessing a certain level of computation capacity.

The defender is assumed to lack access to client training data $\mathcal{D}_{c_i}$, control over the behavior of malicious clients, or even knowledge of the presence and quantity of compromised clients within the FL system.
However, the defender has complete access to model weights $W_{c_i}^r$ and corresponding model updates $\Delta W_{c_i}^r = W_{c_i}^r - G^r$ of client $c_i$ ($c_i \in \mathcal{C}_m^r$) during the $r$-th communication round.
The defender can extract statistical characteristics to identify poisoned weights.

\subsection{Attacker's Goals}

Before elucidating the design objectives of G$^2$uardFL, it is imperative to clarify the objectives of attacker $\mathcal{A}$.

\noindent \textbf{Efficiency}.
The attacker strives to maximize the performance of the poisoned model $F_{\tilde{G}} (\cdot)$ on backdoored samples.

\noindent \textbf{Stealthiness}.
In addition to achieving high performance on backdoored samples, the attacker endeavors to ensure that the performance of the poisoned model $F_{\tilde{G}} (\cdot)$ closely resembles that of the normal model $F_{G} (\cdot)$ on benign samples.
Specifically, the attacker aims to achieve:
$$
\rseq{0.80\linewidth}{!}{
F_{\tilde{G}} (x) = \begin{cases}
    y_c, &\quad \forall x \in \mathcal{D}^{\mathcal{A}}; \\
    F_G (x), &\quad \forall x \in \{\mathcal{D}_{c_i} | c_i \in \mathcal{C} \setminus \mathcal{C}_K \}. \\
\end{cases}
}
$$

The attacker employs various strategies, such as reducing PDR and adjusting loss functions, to minimize the discrepancy between normal weights $W$ and poisoned weights $\tilde{W}$ to a threshold value $\eta$, i.e., $\opn{dist} (W, \tilde{W}) \leq \eta$.
The threshold can be estimated by comparing disparities in distances between local poisoned weights and global weights, as well as the distances between local weights trained on normal data and global weights.

\subsection{Our Design Goals of Defense}\label{subsec:design_goal}

Subsequently, guided by the attacker's objectives, we outline \emph{two-fold} design goals of G$^2$uardFL.

\noindent \textbf{Effectiveness}.
1) After experiencing backdoor attacks, the global model $G^R$ at the final round should perform at a level comparable to a model trained exclusively on clean data.
2) Despite the efficiency of attacks, the defense must ensure that the final global model's performance on backdoored samples is minimized.
3) After poisoned weights from malicious clients impact the global model, the defense should promptly mitigate their influence.

\noindent \textbf{Robustness}.
1) G$^2$uardFL should serve as a universal defense framework capable of countering a wide range of potential backdoor attacks, including data poisoning and model poisoning attacks.
It operates without prior knowledge of specific attack types.
2) The defense should not assume a specific data distribution among clients, whether independent and identically distributed (iid) or non-iid.

\section{G\texorpdfstring{$^2$}uardFL: Overview}\label{sec:overview}

In this section, we present G$^2$uardFL, an effective and robust defense framework designed to mitigate backdoor attacks while preserving the primary task performance.

\noindent \textbf{Motivation}.
G$^2$uardFL redefines the detection of malicious clients by utilizing an \emph{attributed client graph clustering} approach to differentiate model weights originating from malicious clients.
Traditional clustering algorithms, such as K-means and HDBSCAN, rely heavily on the discrimination of model weights and updates when calculating similarities among clients.
In contrast, graph-based clustering methods leverage Graph Convolutional Networks (GCNs) to harness locality for learning more representative and clustering-oriented latent representations.
These representations are then utilized to perform clustering, allowing graph-based methods to achieve improved clustering performance naturally.
Specifically, G$^2$uardFL extracts statistical characteristics from model weights and model updates as client attributes and represents similarities among clients as a graph.

To accurately identify malicious clients, we propose a framework consisting of four key components, as illustrated in Fig.~\ref{fig:overview}.
The complete defense process is presented in Algorithm~\ref{alg:guardfl} in Appendix~\ref{subsec:algorithm}.

\noindent \textbf{Graph Construction}.
This module extracts unique features from client weights and model updates and constructs an attributed graph $\mathcal{G}^r = (\mathcal{C}, \mathbf{E}^r, \mathbf{X}^r)$ to model inter-client relations at the $r$-th communication round.
Here, $\mathbf{E}^r$ and $\mathbf{X}^r$ represent the adjacency matrix and client attributes.
Importantly, the topological structure and client attributes of graph $\mathcal{G}^r$ evolve over communication rounds, rather than remaining static, as highlighted in Subsection~\ref{subsec:threat}.
Further details will be discussed in Subsection~\ref{subsec:graph_construct}.

\noindent \textbf{Graph Clustering}.
This module samples a sub-graph $\hat{\mathcal{G}}^r$ from the entire graph $\mathcal{G}^r$ based on selected clients $\mathcal{C}_m^r$ at the $r$-th round.
Graph-based clustering approaches are then employed to learn more representative latent representations and group clients with similar attributes into the same cluster.
As the orchestration server randomly selects $\mathcal{C}_m^r$ in each round, the number of malicious clients $k$ in $\mathcal{C}_m^r$ also varies, falling within the range of $[0, \min(m, K)]$.
Consequently, the graph-based clustering algorithm must be capable of handling the varying number of clusters.
Further details will be discussed in Subsection~\ref{subsec:graph_clustering}.

\noindent \textbf{Dynamic Filtering and Clipping}.
This module aims to identify both benign and malicious clusters.
Benign clusters (resp. malicious clusters) predominantly consist of benign clients (resp. malicious clients).
The module aggregates the normal and poisoned global models separately from benign and malicious clusters, respectively.
Given that the number of malicious clients $k$ in $\mathcal{C}_m^r$ varies with each round, relying solely on cluster size (i.e., the number of clients per cluster) to identify malicious clusters is not suitable.
Therefore, this module introduces the \emph{benign score} to combine current and historical detection results and utilizes cluster size and benign scores to accurately identify malicious clients.
The positive benign score indicates that the client is consistently assigned to a larger cluster, and its magnitude signifies the corresponding likelihood.
Further details will be discussed in Subsection~\ref{subsec:dynamic_clipping}.

\noindent \textbf{Adaptive Poison Eliminating}.
This module employs an adaptive mechanism to mitigate previously implanted backdoors.
In cases where all selected clients $\mathcal{C}_m^r$ are malicious and the global model is contaminated by poisoned weights, G$^2$uardFL should have a mechanism to purify the global model in subsequent rounds.
After constructing normal and poisoned global models, G$^2$uardFL applies benign scores and geometric properties to adjust the direction of the new global model $G^{r + 1}$, aligning it more closely with normal models and moving it away from poisoned ones.
Further details will be discussed in Subsection~\ref{subsec:adaptive_poison}.

\section{G\texorpdfstring{$^2$}uardFL: Detailed Algorithms}\label{sec:detail}

\subsection{Graph Construction}\label{subsec:graph_construct}

This module is responsible for constructing an attributed graph $\mathcal{G}^r = (\mathcal{C}, \mathbf{E}^r, \mathbf{X}^r)$ to model the inter-client relationships within FL.
The constructed graph is subsequently used to train a model for predicting client behavior.
This module confronts two primary challenges:
1) The first challenge pertains to determining which features should be extracted from the weights and model updates of clients.
These weights and model updates are typically high-dimensional and thus pose difficulties for effective processing directly by clustering algorithms.
2) The second challenge involves constructing the adjacency matrix $\mathbf{E}^r$ in a way that enables accurate differentiation between benign and malicious clients.
This task is challenging because the characteristics of poisoned models generated by different backdoor attacks vary.

\noindent \textbf{Client Attribute Construction}.
To address the first challenge, two types of features are introduced for client attribution representation: \emph{model-wise} features $\mathbf{X}_1$ and \emph{layer-wise} features $\mathbf{X}_2$.
These features are extracted using measures of statistical characteristics $\mathcal{M}$, as detailed in Table~\ref{tab:stats_method}.
Given that these measures compress high-dimensional vectors into a scalar, resulting in significant information loss, we opt for multiple measures to mitigate this concern.
For constructing $\mathbf{X}_1$ features, G$^2$uardFL flattens model weights into a single vector.
However, considering that different layers of a model serve various functions, such as lower layers of convolutional neural networks detecting simple features while higher layers identify more complex concepts, flattening weights from distinct layers into a single vector could lead to the information loss due to the heterogeneous parameter value distribution spaces.
To address this, G$^2$uardFL separately flattens the weights of each layer, calculating layer-wise features $\mathbf{X}_2$ as an additional measure.

Assuming that flattened weights of local model $W_{c_i}^r$ and global model $G^r$ can be presented as $\bs{w}_{c_i}^r$ and $\bs{g}^r$, and that the dimension of model-wise features is $d_1$, model-wise features $\mathbf{X}^r_1 \in \mathbb{R}^{m \times d_1}$ can be formulated as follows:
$$
\rseq{0.88\linewidth}{!}{
    \begin{aligned}
        \mathbf{X}^r_1 = &\left[ m_i(\bs{w}_{c_1}^r), \cdots, m_i(\bs{w}_{c_m}^r)\right]^\top \oplus \left[ m_i(\Delta \bs{w}_{c_1}^r), \cdots, m_i(\Delta \bs{w}_{c_m}^r)\right]^\top \\
        & \oplus \left[ cos(\bs{w}_{c_1}^r, \bs{g}^r), \cdots, cos(\bs{w}_{c_m}^r, \bs{g}^r)\right]^\top \\
    \end{aligned}
},
$$
where the symbol $\oplus$ denotes the operation of horizontally concatenating vectors, $m_i \in \mathcal{M} \setminus \{cos\}$ is a measure of statistical characteristics, and $\Delta \bs{w}_{c_i}^r$ is the flattened vector of model updates $\Delta W_{c_i}^r$, i.e., $\Delta \bs{w}_{c_i}^r = \bs{w}_{c_i}^r - \bs{g}^r$.
This formulation demonstrates model-wise features consist of 19 distinct features (i.e., $d_1 = 19$).

For the computation of layer-wise features $\mathbf{X}_2 \in \mathbb{R}^{m \times d_2}$, weights are first segmented based on the layer they belong to, and then weights of the same layer are integrated into a vector.
Assuming a model $F_{W^r_{c_i}} (\cdot)$ with $L$ layers, weights $W^r_{c_i}$ are partitioned into a set of vectors $\{\bs{w}_{c_i, 1}^r, \bs{w}_{c_i, 2}^r, \cdots, \bs{w}_{c_i, L}^r\}$.
Building on the observation by Zhang \emph{et al.}~\cite{zhang2022fldetector} that model updates from malicious clients are inconsistent in backdoor attacks, we explore changes in model updates between two successive iterations in the layer-wise features.
Specifically, at the $r$-th iteration, we compute model updates $\Delta W_{c_i}^r$ and $\Delta W_{c_i}^{r - 1}$ from the previous iteration of the local client $c_i$.
As $\mathcal{C}_m$ is chosen randomly at each iteration, it is possible that $c_i \in \mathcal{C}_m^r$ is not in $\mathcal{C}_m^{r - 1}$.
In such cases, we replace $W_{c_i}^{r - 1}$ with $G^r$ and set $\Delta W_{c_i}^{r - 1} = G^r - G^{r - 1}$.
The layer-wise features $\mathbf{X}_{2}$ can then be calculated as:
$$
\rseq{0.99\linewidth}{!}{
    \begin{aligned}
        & \mathbf{X}^r_{2} = \left[ m_i(\bs{w}_{c_1, l}^r), \cdots, m_i(\bs{w}_{c_m, l}^r)\right]^\top \oplus \left[ m_i(\Delta \bs{w}_{c_1, l}^r),  \cdots, m_i(\Delta \bs{w}_{c_m, l}^r)\right]^\top \\
        &\; \oplus \left[ cos(\bs{w}_{c_1, l}^r, \bs{g}^r_l), \cdots, cos(\bs{w}_{c_m, l}^r, \bs{g}^r_l)\right]^\top \oplus \left[ cos(\Delta \bs{w}_{c_1, l}^r, \Delta \bs{w}_{c_1, l}^{r - 1}), \cdots, cos(\Delta \bs{w}_{c_m, l}^r, \Delta \bs{w}_{c_m, l}^{r - 1})\right]^\top \\
        &\; \oplus \left[ m_i(\Delta \bs{w}_{c_1, l}^r - \Delta \bs{w}_{c_1, l}^{r - 1}), \cdots, m_i(\Delta \bs{w}_{c_m, l}^r - \Delta \bs{w}_{c_m, l}^{r - 1})\right]^\top \\
    \end{aligned}
},
$$
where $l \in [1, L]$ and $m_i$ has the same meaning with that the above equation.
According to this equation, G$^2$uardFL generates a total of $29L$ (i.e., $d_2 = 29L$) layer-wise features.

G$^2$uardFL leverages Z-Score Normalization~\cite{fei2021score} to normalize features $\mathbf{X}_1^r$ and $\mathbf{X}_2^r$ due to their different magnitudes.
Since the orchestration server randomly selects $K$ clients, G$^2$uardFL utilizes zero vectors to fill client features of non-participating clients, i.e.,
\begin{equation}\label{eq:padding_zero}
\rseq{0.65\linewidth}{!}{
    \mathbf{X}^r_{c_i} = \begin{cases}
        \mathbf{X}^r_{1, c_i} \oplus \mathbf{X}^r_{2, c_i},& \quad c_i \in \mathcal{C}_m^r; \\
        \mymathbb{0}, & \quad c_i \in \mathcal{C} \setminus \mathcal{C}_m^r, \\
    \end{cases}
}
\end{equation}
where $\mathbf{X}^r_{1, c_i}$ and $\mathbf{X}^r_{2, c_i}$ are normalized model-wise and layer-wise features of client $c_i$ at iteration $r$, respectively.

\begin{figure}[t]
    \centering
    \includegraphics[width=0.9\linewidth]{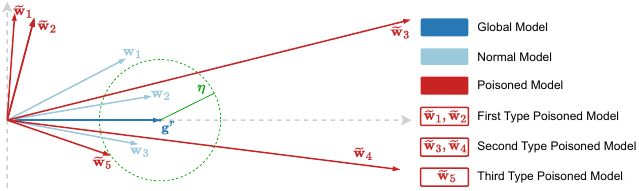}
    \caption{Flattened weights of client models. The two gray dashed lines represent the direction of the global model and its corresponding orthogonal direction, respectively.}\label{fig:flatten_weight}
    \vspace{-10pt}
\end{figure}

\noindent \textbf{Inter-client Relation Construction}.
To address the second challenge, the attention turns towards delineating relationships between normal and poisoned weights, as represented in Fig.~\ref{fig:flatten_weight}.
Weights of benign clients exhibit variations, ascribed to the partially non-iid local data, symbolized by $\bs{w}_1$, $\bs{w}_2$, and $\bs{w}_3$.
Poisoned weights align roughly in the same direction as normal ones to avert degradation of primary task performance, as shown by $\cos(\bs{w}_i, \tilde{\bs{w}}_j) > 0$.

We identify three types of poisoned weights following previous research~\cite{nguyen2022flame}.
The first type, denoted by $\tilde{\bs{w}}_1$ and $\tilde{\bs{w}}_2$, demonstrates a large angular deviation from local models of benign clients and the global model.
These weights are trained to optimize performance on backdoored samples, either by high PDR or large local training rounds.
The second type, denoted by $\tilde{\bs{w}}_3$ and $\tilde{\bs{w}}_4$, possesses a small deviation but exhibits a large magnitude intended to amplify the impact of the attacker.
This magnitude is a product of scaling up weights to augment their influence on the global model~\cite{wang2020attack}.
In the third category of attacks, the angular and magnitude variations are not substantially different from normal models (i.e., $\tilde{\bs{w}}_5$), rendering them less discernible from normal models.
These poisoned weights are crafted by constraining the $L_p$-norm of the difference between poisoned weights and the global model, as seen in the Constrain-and-scale attack~\cite{bagdasaryan2020how} and AnaFL~\cite{bhagoji2019analyzing}.
Next, we develop a method for assessing inter-client relationships by reducing similarities between benign and malicious clients, while simultaneously enhancing similarities among malicious clients.

For the first type, where poisoned weights consistently maintain a large angle with normal ones, G$^2$uardFL applies cosine similarity to measure distances among clients.
The adjacency matrix $\mathbf{E}^r_1$ is thus formulated as follows:
\begin{equation}\label{eq:cos_edge}
\rseq{0.80\linewidth}{!}{
    \mathbf{E}^r_1 = \left[
        \begin{matrix}
        1 & \frac{1 + \opn{cos}(\bs{w}_{c_1}^r, \bs{w}_{c_2}^r) }{2} & \cdots & \frac{1 + \opn{cos}(\bs{w}_{c_1}^r, \bs{w}_{c_m}^r) }{2} \\
        \frac{1 + \opn{cos}(\bs{w}_{c_2}^r, \bs{w}_{c_1}^r) }{2} & 1 & \cdots & \frac{1 + \opn{cos}(\bs{w}_{c_2}^r, \bs{w}_{c_m}^r) }{2} \\
        \vdots & \vdots & \ddots & \vdots \\
        \frac{1 + \opn{cos}(\bs{w}_{c_m}^r, \bs{w}_{c_1}^r) }{2} & \frac{1 + \opn{cos}(\bs{w}_{c_m}^r, \bs{w}_{c_2}^r)}{2} & \cdots & 1 \\
        \end{matrix}
        \right]
}.
\end{equation}

To measure the magnitude difference between weights for the second type of attack, G$^2$uardFL substitutes the $\nicefrac{[1 + \opn{cos}(\bs{w}_{c_i}^r, \bs{w}_{c_j}^r)]}{2}$ term in Eq.~\eqref{eq:cos_edge} with $\opn{abs}(\lVert \bs{w}_{c_i}^r \rVert_2 - \lVert \bs{w}_{c_j}^r \rVert_2)$ to construct the adjacency matrix $\mathbf{E}^r_2$.
The rationale is that the magnitude of poisoned weights consistently exceeds that of normal weights.
Consequently, when flattened weights of normal and poisoned models have a small deviation, model differences between normal and poisoned models are of a larger magnitude than those between normal models.

To enhance the similarity between malicious clients for the third type of backdoor attacks, G$^2$uardFL computes the difference between $L_2$ norms of model updates of two clients $\opn{abs}(\lVert \Delta \bs{w}_{c_i}^r \rVert_2 - \lVert \Delta \bs{w}_{c_j}^r \rVert_2)$ and uses this to construct adjacency matrix $\mathbf{E}^r_3$.
This stems from the fact that this type of attack constrains the magnitude of model updates of malicious clients to a pre-defined small constant $\eta$, as shown by the circle enclosed by a dashed line in Fig.~\ref{fig:flatten_weight}, resulting in an absolute difference between poisoned weights that is close to $0$.

Given that the defender is unaware of the type of backdoor attacks, G$^2$uardFL applies the Z-Score Normalization to normalize each adjacency matrix and integrates these three adjacency matrices.
Note that for the $L_p$ norm-based matrices, a smaller norm value signifies a higher similarity, which is the inverse of the cosine similarity-based matrix.
Consequently, G$^2$uardFL alters elements within $\mathbf{E}_2^r$ and $\mathbf{E}_3^r$ to the opposite number after the normalization of norm-based matrices.

G$^2$uardFL focuses exclusively on positive inter-client relations, considering elements of adjacency matrices below $0$ as non-existent edges, and scales the intensity of relations between clients to $[0, 1]$, i.e.,
$$
\rseq{0.92\linewidth}{!}{
    \mathbf{E}_i^r = \opn{exp} (- \opn{tanh}(\opn{clip}_{min} (\mathbf{E}_i^r, 0))), \quad \forall i \in \{1, 2, 3\}
}.
$$
Finally, G$^2$uardFL sets relations of non-participating clients to $0$ and constructs the adjacency matrix $\mathbf{E}^r$ at iteration $r$, i.e.,
$$
\rseq{0.88\linewidth}{!}{
    \mathbf{E}^r_{c_i, c_j} =
    \begin{cases}
        0, &c_i \notin \mathcal{C}_m^r \vee c_j \notin \mathcal{C}_m^r; \\
        \nicefrac{1}{3} \sum_{i = 1}^3 \mathbf{E}^r_{i, c_i, c_j},  &c_i \in \mathcal{C}_m^r \wedge c_j \in \mathcal{C}_m^r.\\
    \end{cases}
}
$$

The graph construction method illustrated above constructs an attributed graph at each round.
This could result in substantial differences in features of the same client or inter-client relations of client pairs between two successive rounds.
To mitigate this issue, G$^2$uardFL introduces a momentum component to smooth the attributed graph as follows:
\begin{equation}\label{eq:smooth_graph}
\rseq{0.50\linewidth}{!}{
    \begin{aligned}
        \mathbf{X}^r &= (1 - \kappa_1) \mathbf{X}^{r - 1} + \kappa_1 \mathbf{X}^r; \\
        \mathbf{E}^r &= (1 - \kappa_2) \mathbf{E}^{r - 1} + \kappa_2 \mathbf{E}^r, \\
    \end{aligned}
}
\end{equation}
where $\kappa_1$ and $\kappa_2$, both within the range [0, 1], are employed to balance historical and current graphs.

\subsection{Graph Clustering}\label{subsec:graph_clustering}

\begin{figure}[t]
    \centering
    \includegraphics[width=1.0\linewidth]{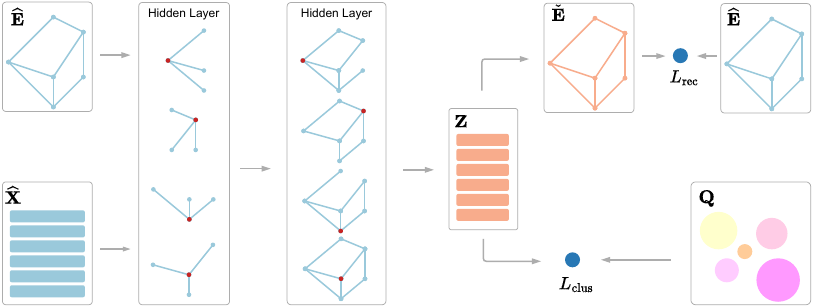}
    \caption{The conceptual design for GAE-based clustering.}\label{fig:graph_clustering}
    \vspace{-10pt}
\end{figure}

This module leverages graph clustering techniques to segregate clients with similar attributes into corresponding clusters.
Compared with traditional clustering algorithms (e.g., K-means and HBDSCAN), graph-based clustering methods utilize GCNs to learn more representative representations from client attributes and use these representations to perform clustering instead of directly using client attributes.
Thus, graph-based clustering methods can naturally achieve better clustering performance.
A challenge arises when considering the indeterminate quantity of clusters since $m$ clients are arbitrarily chosen by server $\mathcal{S}$ at each iteration.

\noindent \textbf{Clustering Design}.
The preceding module produces an attributed graph $\mathcal{G}^r = (\mathcal{C}, \mathbf{E}^r, \mathbf{X}^r)$.
G$^2$uardFL applies Eq.~\eqref{eq:sample_graph} to extract an attributed sub-graph $\hat{\mathcal{G}}^r = (\mathcal{C}_m, \hat{\mathbf{E}}^r, \hat{\mathbf{X}}^r)$, which comprises $m$ clients selected by the server at each round.
\begin{equation}\label{eq:sample_graph}
\rseq{0.88\linewidth}{!}{
    \hat{\mathbf{E}}^r = \left\{ \mathbf{E}^r_{c_i, c_j} | c_i \in \mathcal{C}_m \wedge c_j \in \mathcal{C}_m \right\},  \;
    \hat{\mathbf{X}}^r = \left\{ \mathbf{X}^r_{c_i} | c_i \in \mathcal{C}_m \right\}
}.
\end{equation}

According to the framework employed by recent graph clustering methods~\cite{hui2020collaborative,mrabah2022rethinking,pan2018adversarially}, G$^2$uardFL employs Graph Auto-Encoders (GAEs) to transform client attributes into lower-dimension space (i.e., latent representations) via GCNs.
These representations are then used for adjacency matrix reconstruction and clustering assignments, as illustrated in Fig.~\ref{fig:graph_clustering}.
Specifically, G$^2$uardFL integrates a self-supervision task (i.e., reconstruction) to obtain clustering-oriented latent representations, and implements a pseudo-supervision task (i.e., clustering) to segregate clients into distinct clusters.

Initially, this module uses a non-linear encoder $\opn{Enc} (\cdot, \cdot)$, based on GCNs, to craft low-dimensional latent representations depicted by the matrix $\mathbf{Z} \in \mathbb{R}^{m \times d_3}$.
\begin{equation}\label{eq:graph_encode}
    \mathbf{Z} = \opn{Enc}(\hat{\mathbf{X}}^r, \hat{\mathbf{E}}^r).
\end{equation}
Here, $d_3$ represents the dimensionality of latent representations.
Later, in the reconstruction task, this module applies derived features $\mathbf{Z}$ to reconstruct the adjacency matrix $\hat{\mathbf{E}}^r$.
\begin{equation}\label{eq:edge_reconstruction}
    \check{\mathbf{E}}^r = \opn{Sigmoid}(\mathbf{Z} \times \mathbf{Z}^\top),
\end{equation}
where $\opn{Sigmoid} (\cdot)$ limits the magnitude of elements within the reconstructed adjacency matrix $\check{\mathbf{E}}^r$ to $[0, 1]^{m \times m}$.
Assuming the existence of $Q$ clusters among clients, the module uses the matrix $\mathbf{Q} \in \mathbb{R}^{Q \times d_3}$ to represent clustering centroids.
To address the challenge initially mentioned, Eq.~\eqref{eq:clustering} calculates the similarity between latent representations and clustering centers.
\begin{equation}\label{eq:clustering}
\rseq{0.72\linewidth}{!}{
    \mathbf{P}_{i, j} = \frac{\opn{exp} \left( - \frac{1}{2} \left( \mathbf{Z}_i - \mathbf{Q}_j \right)^\top \left( \mathbf{Z}_i - \mathbf{Q}_j \right) \right)}{\sum_{q = 1}^Q \opn{exp} \left( - \frac{1}{2} \left( \mathbf{Z}_i - \mathbf{Q}_q \right)^\top \left( \mathbf{Z}_i - \mathbf{Q}_q \right) \right)}
}
,
\end{equation}
where $\mathbf{P} \in [0, 1]^{m \times Q}$ is the soft clustering assignment matrix, $\mathbf{Z}_i$ is the $i$-th row of latent presentations $\mathbf{Z}$, and $\mathbf{Q}_j$ is the $j$-th row of the matrix $\mathbf{Q}$ (i.e., the $j$-th clustering center).
The cluster to which the $i$-th client of chosen clients $\mathcal{C}_m$ belongs is determined by $p_i = \opn{arg \; max}_{j \in \{1, \cdots, Q\}} (\mathbf{P}_{i, j})$.
The hard clustering assignment can be formulated as follows:
\begin{equation}\label{eq:hard_assignment}
    \hat{\mathbf{P}}_{i, j} = \begin{cases}
        1, \quad j = p_i; \\
        0, \quad j \neq p_i.
    \end{cases}
\end{equation}
Note that, while $Q$ is pre-defined, Eq.~\eqref{eq:hard_assignment} shows the possibility of certain clusters having no clients.

\noindent \textbf{Loss Functions}.
For the reconstruction task, a loss function $L_{\opn{rec}}$ is introduced to minimize the discrepancy between the original adjacency matrix $\hat{\mathbf{E}}^r$ and the reconstructed one $\check{\mathbf{E}}^r$:
\begin{equation}\label{eq:loss_recon}
\rseq{0.83\linewidth}{!}{
    L_{\opn{rec}} = - \sum_{i, j} \left( \hat{\mathbf{E}}_{i, j}^r \opn{log} (\check{\mathbf{E}}_{i, j}^r) + (1 - \hat{\mathbf{E}}_{i, j}^r) \opn{log} (1 - \check{\mathbf{E}}_{i, j}^r) \right)
}.
\end{equation}
In Eq.~\eqref{eq:loss_recon}, the ground-truth adjacency weight $\check{\mathbf{E}}_{i, j}^r \in [0, 1]$ can be considered as a soft label.
The clustering task loss is represented by the Kullback-Leibler divergence between the soft clustering assignment $\mathbf{P}$ and its corresponding hard clustering assignment $\hat{\mathbf{P}}$:
\begin{equation}\label{eq:loss_cluster}
\rseq{0.76\linewidth}{!}{
    L_{\opn{clus}} = \opn{KL} (\hat{\mathbf{P}} \Vert \mathbf{P}) = \sum_{i, j} \hat{\mathbf{P}}_{i, j} \opn{log} \left( \nicefrac{\hat{\mathbf{P}}_{i, j}}{\mathbf{P}_{i, j}} \right)
}.
\end{equation}
To achieve a balance between these two loss functions, the module incorporates a positive, small constant $\lambda \in [0, 1]$ to formulate the total loss function:
\begin{equation}\label{eq:total_loss}
    L = L_{\opn{rec}} + \lambda L_{\opn{clus}}.
\end{equation}

To obtain clustering-oriented latent representations, G$^2$uardFL first pre-trains the weights of $\opn{Enc}(\cdot)$ solely using the loss function $L_{\opn{rec}}$ for multiple iterations.
Subsequently, the model is trained to optimize both the latent representations and clustering centers by minimizing $L$.
Before training, the clustering center matrix $\mathbf{Q}$ is initialized using clustering centers generated by K-means, with the number of clusters being initialized by HBDSCAN.
Throughout the training, we employ $L_{\opn{clus}}$ to optimize the matrix $\mathbf{Q}$.

\subsection{Dynamic Filtering and Clipping}\label{subsec:dynamic_clipping}

Once clusters have been established in the preceding module, it is challenging to determine which clusters largely consist of malicious clients.
As discussed in Section~\ref{sec:threat_and_design}, we anticipate that less than 50\% of all clients will be malicious, thereby constituting a minority in most instances.
However, the precise number of malicious clients in $\mathcal{C}_m^r$ remains undefined.
Therefore, this module introduces the benign score, which combines historical and current clustering results, and simultaneously utilizes both the cluster size and benign scores to pinpoint malicious clusters.
The rationale of benign scores is that, although poisoned models may dominate in specific rounds, they constitute a minority on average.

\noindent \textbf{Cluster Filtering}.
This module assigns a \emph{benign score} $s_{c_i} \in \mymathbb{R}$ to each client $c_i \in \mathcal{C}$.
A vector $\mathbf{s}_m^r$ is produced to symbolize benign scores of clients within $\mathcal{C}_m^r$ at the $r$-th round.
As a result, the score of a cluster being benign can be calculated as follows:
\begin{equation}\label{eq:cluster_filtering}
\rseq{0.85\linewidth}{!}{
    \bs{p} = \nicefrac{\kappa_3}{m} \left( \hat{\mathbf{P}}^\top \times \mymathbb{1} \right) + \left(\hat{\mathbf{P}}^\top \times \mathbf{s}^r_m \right) \circ \left( \hat{\mathbf{P}}^\top \times \mymathbb{1} \right)^{\circ -1}
},
\end{equation}
where $\mymathbb{1}$ is a vector with all elements equal to $1$.
The Hadamard product and Hadamard inverse are denoted by $\circ$ and ${}^{\circ -1}$, respectively~\cite{reams1999hadamard}.
The first term signifies the ratio between the cluster size and the number of chosen clients per round, while the second term computes the average benign score for each cluster.
Next, we can obtain benign cluster $p^{+} = \opn{arg \; max}_{j \in \{1, \cdots, Q\}} \; \bs{p}_j$ and malicious cluster $p^{-} = \opn{arg \; min}_{j \in \{1, \cdots, Q\}} \; \bs{p}_j$, and further acquire clients within clusters $p^{+}$ and $p^{-}$ as follows:
\begin{equation}\label{eq:pick_up_client}
\rseq{0.55\linewidth}{!}{
    \begin{aligned}
        \hat{\mathcal{C}}^{+} &= \left\{ c_i | c_i \in \mathcal{C}_m^r \wedge \hat{\mathbf{P}}_{i, p^{+}} = 1 \right\}; \\
        \mathcal{C}^{-} &= \left\{ c_i | c_i \in \mathcal{C}_m^r \wedge \hat{\mathbf{P}}_{i, p^{-}} = 1 \right\},
    \end{aligned}
}
\end{equation}
where $\hat{\mathcal{C}}^{+}$ and $\mathcal{C}^{-}$ are considered to be clusters where benign and malicious clients dominate, respectively.

However, as Eq.~\eqref{eq:cluster_filtering} utilizes the average benign score of each cluster to pinpoint malicious clients, there might be scenarios where some clients within the benign cluster $p^{+}$ have low benign scores.
To address this concern, this module utilizes statistical data of benign scores within the benign cluster $p^{+}$ and calculates the distance between the latent representations of clients and their respective cluster centers.
Based on this information, this module identifies and excludes clients that do not meet the criteria for inclusion.
This module calculates the distance using Eq.~\eqref{eq:distance_to_center}:
\begin{equation}\label{eq:distance_to_center}
\rseq{0.34\linewidth}{!}{
    \bs{d} = \left\Vert \mathbf{Z} - \hat{\mathbf{P}} \times \mathbf{Q} \right\Vert_2
}.
\end{equation}
Next, this module utilizes the $\alpha_1$ percentile $P_{\alpha_1} (\cdot)$ of benign scores and the $\alpha_2$ percentile $P_{\alpha_2} (\cdot)$ of distances within the benign cluster $p^{+}$ to filter out clients as follows:
\begin{equation}\label{eq:filter_client}
\rseq{0.86\linewidth}{!}{
    \mathcal{C}^{+} = \left\{ c_i | c_i \in \hat{\mathcal{C}}^{+} \wedge \bs{s}^r_{m, i} \geq \opn{P_{\alpha_1}}(\bs{s}^r_m)  \wedge \bs{d}^+_i \leq \opn{P_{\alpha_2}}(\bs{d^+}) \right\}
},
\end{equation}
where $\bs{d}^+ = \{ \bs{d}_i | c_i \in \hat{\mathcal{C}}^{+} \}$ indicates the vector of distances of clients within benign cluster $p^+$.

Having distinguished benign clients $\mathcal{C}^{+}$ and malicious clients $\mathcal{C}^{-}$, benign scores should be further updated using
\begin{equation}\label{eq:benign_score_update}
\rseq{0.88\linewidth}{!}{
    \bs{s}_{c_i} = \begin{cases}
    \bs{s}_{c_i} + \kappa_4 \opn{abs}(\bs{s}_{c_i}) \mathbf{P}_{c_i, p^{+}}, \quad c_i \in \mathcal{C}^{+}; \\
    \bs{s}_{c_i} - \kappa_4 \opn{abs}(\bs{s}_{c_i}), \qquad \quad \ \ c_i \in \mathcal{C}^{-}; \\
    \bs{s}_{c_i}, \qquad \qquad \qquad \qquad \quad \: \ c_i \notin \mathcal{C}^{+} \wedge c_i \notin \mathcal{C}^{-}, \\
    \end{cases}
}
\end{equation}
where $\kappa_4 \in (0, 1]$ is the factor to balance historical and current benign scores.
Examining Eq.~\eqref{eq:benign_score_update}, the magnitude of benign scores indicates the likelihood of a client being assigned to a cluster with a larger or smaller size.
Finally, benign scores should be normalized to attribute negative values to malicious clients, hence ensuring that such clients exert a detrimental influence, represented by $\bs{s}^{r + 1} = \opn{tanh} \left( \bs{s}^r \right)$.
Furthermore, benign scores are randomly initialized in the first round, enabling G$^2$uardFL to refine and update benign scores based on the clustering results.

\noindent \textbf{Norm Clipping}.
As shown in Fig.~\ref{fig:flatten_weight}, poisoned weights (e.g., $\tilde{\bs{w}}_1$, $\tilde{\bs{w}}_2$, $\tilde{\bs{w}}_3$, and $\tilde{\bs{w}}_4$) derived from the preceding two types of backdoor attacks demonstrate a significant divergence from the global model.
When these poisoned weights are inadvertently aggregated into the global model, the corrupted global model deviates from benign clients abruptly, thereby leading to an unstable training process.
To minimize the impact of these poisoned models, this module employs the average median norm of model updates to restrict this divergence.

This module computes the average median norm $n^r$ of model updates at the $r$-th round as follows:
\begin{equation}\label{eq:average_median_norm}
\rseq{0.86\linewidth}{!}{
    n^r = n^{r - 1} + \nicefrac{1}{r} \left( \opn{median} (\left\Vert \Delta W_{c_0}^r \right\Vert_2, \cdots, \left\Vert \Delta W_{c_m}^r \right\Vert_2) - n^{r - 1} \right)
}, 
\end{equation}
where $\Vert \Delta W_{c_i}^r \Vert_2$ is the $L_2$ norm of model update of the client $c_i \in \mathcal{C}_m^r$.
This module then limits the norm of weights of benign and malicious clients to be lower than $n^r$:
\begin{equation}\label{eq:limit_norm}
\rseq{0.72\linewidth}{!}{
    \overline{W}_{c_i}^r = G^{r - 1} + \Delta W_{c_i}^r \cdot \opn{min} (1.0, \nicefrac{n^r}{ \left \Vert \Delta W_{c_i}^r \right \Vert_2} )
}
.
\end{equation}
Note that norm clipping is executed on every client chosen in the $r$-th round.
This procedure aids in mitigating the influence of malicious clients on the global model, thereby enhancing the robustness of the FL process.

\subsection{Adaptive Poison Eliminating}\label{subsec:adaptive_poison}

At the end of each FL round, the orchestration server performs the aggregation of model weights from participating clients to form a new global model.
However, this process can potentially introduce poisoned weights into the global model, thus inadvertently implanting backdoors.
To mitigate this risk, we introduce an adaptive mechanism within this module, which independently aggregates normal and poisoned global models and adjusts the direction of the new global model away from poisoned models.

The module separately aggregates normal and poisoned global models, as detailed in Eq.~\eqref{eq:aggregate_benign_model}:
\begin{equation}\label{eq:aggregate_benign_model}
\rseq{0.80\linewidth}{!}{
    \begin{aligned}
        G^{+} &= \sum_{c_i \in \mathcal{C}^+} G^{r - 1} + \left( \overline{W}_{c_i}^r - G^{r - 1} \right) \cdot \nicefrac{e^{\bs{d}^{+}_i}}{\sum_{j} e^{\bs{d}^{+}_j}}; \\
        G^{-} &= \sum_{c_i \in \mathcal{C}^-} G^{r - 1} + \left( \overline{W}_{c_i}^r - G^{r - 1} \right) \cdot \nicefrac{e^{\bs{d}^{-}_i}}{\sum_{j} e^{\bs{d}^{-}_j}},
    \end{aligned}
}
\end{equation}
where $\bs{d}^+ = \left\{ \bs{d}_i | c_i \in \mathcal{C}^+ \right\}$ and $\bs{d}^- = \left\{ \bs{d}_i | c_i \in \mathcal{C}^- \right\}$ represent distance vectors between latent representations of clients and their corresponding cluster centers.
Unlike FedAvg, which assumes importance according to the size of local data for each client, Eq.~\eqref{eq:aggregate_benign_model} considers that client weights closer to cluster centers should be given heightened attention.

As previously mentioned, the global model $G^{r - 1}$ may have backdoors introduced in previous rounds.
To reduce the impact of these backdoors on the global model, and to impair the success of backdoor attacks, G$^2$uardFL strives to widen differences between new global model $G^r$ and poisoned global model $G^{-}$.
In practice, given that poisoned weights in the second type of backdoor attacks show significant deviation from the global model, this module limits the magnitude of model updates between $G^r$ and $G^{-}$ to the same level as the magnitude of model updates between $G^r$ and $G^{+}$:
\begin{equation}\label{eq:limit_malicious_global}
\rseq{0.84\linewidth}{!}{
    \overline{G}^{-} = G^{r - 1} + \left( G^{-} - G^{r - 1}\right) \cdot \opn{min} (1.0, \frac{\left \Vert G^{+} - G^{r - 1}  \right \Vert_2}{\left \Vert G^{-} - G^{r - 1}  \right \Vert_2})
}.
\end{equation}
Subsequently, this module increases the disparity between normal and poisoned global models to aggregate the new global model:
\begin{equation}\label{eq:aggregate_global}
\rseq{0.84\linewidth}{!}{
    G^{r} = G^{+} + \gamma \frac{\sum_{c_j \in \mathcal{C}^-} \opn{abs} (\bs{s}_{m, j}^r)}{\sum_{c_j \in \mathcal{C}_m^r} \opn{abs}(\bs{s}_{m, j}^r)} \cdot \opn{log} \left( 1.0 + n^r \right) \cdot \left( G^+ - \overline{G}^- \right)
},
\end{equation}
where $\gamma$ is a positive number and set to $0.01$ in experiments.

Eq.~\eqref{eq:aggregate_global} considers that the magnitude of the divergence between normal and poisoned global models is influenced by three factors: the relative level of maliciousness of clients within a malicious cluster, the average median norm of model updates, and the difference between two global models.
When the poisoned global model $G^{-}$ significantly diverges from normal global model $G^{+}$, the aggregated new global model $G^{r + 1}$ will veers further away from $G^{-}$.
Consequently, the performance of $G^{r + 1}$ on backdoored samples diminishes, compelling malicious clients to generate larger model updates in an attempt to alter the update direction of the global model towards backdoor implantation, thereby increasing the chance of detection.

In summary, we design the first three modules for constructing the client attribute graph and utilizing the graph clustering to precisely identify malicious clients, and introduce the fourth module for eliminating the influence of previously embedded backdoors.

\section{Convergence Analysis}\label{sec:convergence_analysis}

In this section, we delve into the convergence analysis of our proposed G$^2$uardFL.
We commence by establishing certain assumptions regarding functions $\{ F_{c_i} (\cdot) \vert c_i \in \mathcal{C} \}$, similar to those commonly made in previous research works~\cite{fourati2023filfl,li2020convergence,mirzasoleiman2020coresets,wang2022quantized}.

\begin{assumption}\label{assump:smooth}
Each local function $F_{c_i} (\cdot)$ is lower-bounded, i.e., $F_{c_i} (W) \geq F(G^\star) \geq - \infty$ and is $L$-smooth: for all $V$ and $W$,
$$
F_{c_i} (V) \leq F_{c_i} (W) + (V - W)^\top \nabla F_{c_i} (W) + \frac{L}{2} \left \lVert V - W \right \rVert_2^2.
$$
\end{assumption}

\begin{assumption}\label{assump:norm_variance}
Let $\xi_{c_i}$ be randomly sampled from the local data of client $c_i$.
The mean of stochastic gradients is unbiased:
$$
\mymathbb{E} \left[ \nabla F_{c_i} (W_{c_i}; \xi_{c_i}) \right] = \mymathbb{E} \left[ \nabla F_{c_i} (W_{c_i}) \right].
$$
Moreover, the variance of stochastic gradients is bounded:
$$
\mymathbb{E} \left[ \lVert \nabla F_{c_i} (W_{c_i}; \xi_{c_i}) - \nabla F_{c_i} (W_{c_i}) \rVert^2 \right] \leq \zeta^2.
$$
\end{assumption}

\begin{assumption}\label{assump:norm_bounded}
We assume there exist $M$ constants $D_i \geq 0$ such that:
$$
\mymathbb{E} \left[ \lVert \nabla F_{c_i} (W) - \nabla F(W) \rVert^2 \right] \leq D_i^2, \quad \forall i = 1, \cdots, M.
$$
These $D_i$ quantify the heterogeneity of local datasets.
\end{assumption}

With these assumptions in place, we proceed with the convergence analysis.
It is assumed that each client conducts $E$ local stochastic gradient descent updates per FL round.
The FL process concludes after $R$ rounds, yielding $G^R$ as the final global model. 
The subsequent theorem articulates the convergence properties of G$^2$uardFL.

\begin{theorem}\label{thm:converage}
Under Assumptions \ref{assump:smooth} to \ref{assump:norm_bounded}, where $L$, $\zeta$ and $D_i$ are defined as previously described, if we choose $\lambda=\nicefrac{K^{\frac{1}{2}}}{8 L T^{\frac{1}{2}}}$ and $E \leq \nicefrac{T^{\frac{1}{4}}}{K^{\frac{3}{4}}}$ where $T = E R \geq \max \left\{16 K^3, \frac{16}{K}, \frac{16}{L^2}\right\}$ denotes the total number of stochastic gradient descent updates per client, then we have:
\begin{equation}
\rseq{0.78\linewidth}{!}{
\begin{aligned}
    & \frac{1}{R} \sum_{r = 1}^R \mymathbb{E} \left[ \lVert \nabla F(G^{r - 1}) \rVert^2 \right] \leq \frac{20 L \left(\mathbb{E}[F(G^0)]-F(G^{\star}) \right)}{T^{\frac{1}{2}} K^{\frac{3}{2}}}+ \\
    &\quad \frac{21 K + 11}{4 K} \sum_{c_i \in \mathcal{C}^{+}} D_i^2+\frac{K + 1}{4 K} \sum_{c_i \in \mathcal{C}^{-}} D_i^2+\frac{1}{4} \zeta^2
\end{aligned}
},
\end{equation}
where $K$ represents the number of chosen clients in each iteration and $\lambda$ denotes the learning rate.
\end{theorem}

The proof of this theorem is provided in Appendix~\ref{sec:proof}.
Additionally, since we assume $\gamma = \frac{\lambda}{C\sqrt{K}}$ with $C > 0$ being the constant in this proof, it is evident that the update direction alterations introduced by the adaptive poison elimination module in Subsection~\ref{subsec:adaptive_poison} will not impede the convergence of the global model when a small positive constant is assigned to $\gamma$, alongside the utilization of a decaying learning rate.
The ablation study of this module can be found in Appendix.~\ref{subsec:ada_impact}.

\section{Experimental Evaluation}\label{sec:experiment}

In this section, we conduct comprehensive evaluations of G$^2$uardFL's defense abilities against backdoor attacks.
Specifically, we seek to answer the following questions:
\begin{itemize}[leftmargin=10pt]
    \item Can G$^2$uardFL outperform existing state-of-the-art defenses in terms of both attack effectiveness and primary task performance? (Refer to Subsection~\ref{subsec:evaluate_effectiveness})
    \item Can G$^2$uardFL demonstrate sufficient robustness against various backdoor attacks, accommodate data heterogeneity, and adaptive attacks? (Refer to Subsection~\ref{subsec:evaluate_robustness})
    \item Can G$^2$uardFL maintain stable performance regardless of variations in hyper-parameters, such as the number of malicious clients or the number of total clients per round? (Refer to Subsection~\ref{subsec:evaluate_sensitivity})
\end{itemize}

\begin{table*}[t]
    \centering
    \caption{Effectiveness of G$^2$uardFL in comparison to state-of-the-art defenses for backdoor attacks(\%). 
    We select \emph{3DFed} attack~\cite{li20233dfed} for MNIST and CIFAR-10, and \emph{PGD with replacement} attack~\cite{wang2020attack} for Sentiment-140 and Reddit.}\label{tab:effective_result}
    \renewcommand\arraystretch{1.5}
    \scalebox{0.62}{
    \begin{threeparttable}
        \begin{tabular}{p{2.2cm} p{1.6cm}<{\centering} p{1.6cm}<{\centering} p{1.6cm}<{\centering} p{1.6cm}<{\centering} p{1.6cm}<{\centering} p{1.6cm}<{\centering} p{1.6cm}<{\centering} p{1.6cm}<{\centering} p{1.6cm}<{\centering} p{1.6cm}<{\centering} p{1.6cm}<{\centering} p{1.6cm}<{\centering}}
            \toprule[1.2pt]
            \multirow{2}{*}{\parbox{2.5cm}{\centering Defense}} & \multicolumn{3}{c}{MNIST} & \multicolumn{3}{c}{CIFAR-10} & \multicolumn{3}{c}{Sentiment-140} & \multicolumn{3}{c}{Reddit} \\
            ~ & ASR & ACC & DS & ASR & ACC & DS & ASR & ACC & DS & ASR & ACC & DS \\
            \midrule[1pt]
            \emph{Benign setting} & 09.64 & 99.20 & - & 09.93 & 73.71 & - & 19.80 & 76.59 & - & 00.00 & 22.67 & - \\
            \emph{No defense} & 98.18 & 97.52 & - & 96.16 & 73.08 & - & 100.00 & 64.51 & - & 100.00 & 22.53 & - \\
            \midrule[1pt]
            FLAME~\cite{nguyen2022flame} & 87.22 & 98.91 & 22.64 & 42.34 & \textbf{73.86} & 64.76 & 99.01 & 65.82 & 1.95 & 99.49 & 22.52 & 1.00 \\
            DeepSight~\cite{rieger2022deep} & 98.86 & 99.04 & 02.25 & 73.01 & 73.80 & 39.52 & 100.00 & 50.00 & 0.00 & OOM & OOM & OOM \\
            FoolsGold~\cite{fung2020limitations} & 20.68 & 98.83 & 88.01 & 98.05 & 73.36 & 03.80 & 99.01 & 66.13 & 1.95 & 100.00 & 22.17 & 0.00 \\
            Krum~\cite{blanchard2017machine}  & 93.30 & 93.97 & 01.39 & 88.19 & 72.51 & 20.31 & 94.06 & 67.90 & 10.92 & 75.26 & 22.68 & 23.67 \\
            Multi-Krum~\cite{blanchard2017machine}  & 12.36 & 96.24 & 91.74 & 50.27 & 73.54 & 59.34 & 100.00 & 65.10 & 0.00 & 100.00 & 22.64 & 0.00 \\
            RLR~\cite{ozdayi2021defending} & 40.83 & 98.43 & 73.91 & 23.32 & 72.51 & 74.54 & 27.72 & \textbf{75.44} & 73.58 & 35.79 & \textbf{22.81} & 33.66 \\
            FLTrust~\cite{cao2021fltrust} & 18.07 & 98.75 & 89.56 & 23.28 & 72.51 & 74.54 & 51.49 & 70.04 & 57.32 & 100.00 & 22.69 & 0.00 \\
            NDC~\cite{sun2019can} & 95.13 & \textbf{99.19} & 09.28 & 24.48 & 72.32 & 73.89 & 100.00 & 64.67 & 0.00 & 100.00 & 22.62 & 0.00 \\
            RFA~\cite{pillutla2019robust} & 12.27 & 96.32 & 91.82 & 19.54 & 73.58 & 76.87 & 100.00 & 65.36 & 0.00 & 100.00 & 22.64 & 0.00 \\
            Weak DP~\cite{dwork2014algorithmic} & 94.67 & 99.05 & 10.12 & 22.48 & 72.58 & 74.97 & 100.00 & 64.30 & 0.00 & 100.00 & 20.81 & 0.00 \\
            G$^2$uardFL (\textbf{ours}) & \textbf{09.98} {\scriptsize $\downarrow 2.29$} & 97.84 {\scriptsize $\downarrow 1.35$} & \textbf{93.77} {\scriptsize $\uparrow 1.95$} & \textbf{10.61} {\scriptsize $\downarrow 8.93$} & 73.05  {\scriptsize $\downarrow 0.81$} & \textbf{80.40} {\scriptsize $\uparrow 5.78$} & \textbf{7.92}  {\scriptsize $\downarrow 19.80$} & 74.12  {\scriptsize $\downarrow 1.32$}  & \textbf{82.13} {\scriptsize $\uparrow 8.55$} & \textbf{0.00}  {\scriptsize $\downarrow 35.79$} & 22.48  {\scriptsize $\downarrow 0.33$} & \textbf{36.71} {\scriptsize $\uparrow 3.05$} \\
            \bottomrule[1.2pt]
        \end{tabular}
    \end{threeparttable}
    }
    \vspace{-10pt}
\end{table*}

\subsection{Experimental Setup}\label{subsec:experimental_setup}

\noindent \textbf{Implementation}.
We have implemented G$^2$uardFL in PyTorch~\cite{paszke2019pytorch}.
Our simulated FL system follows the established framework used in previous works~\cite{wang2020attack,mcmahan2017communication,bagdasaryan2020how}.
In each communication round, the orchestrating server randomly picks a subset of clients and shares the current global model with them.
Selected clients perform local training for $E$ stochastic gradient descent steps and transmit the updated model weights to the server.
Our FL system is deployed on a server equipped with 64GB RAM, an Intel(R) Core(TM) I9-10900X@3.70GHz CPU, and an Nvidia RTX 3080 GPU.

For the non-linear encoder $\opn{Enc}(\cdot, \cdot)$, we employ two GCN layers~\cite{kipf2017semi} as the backbone for graph embedding.
The dimension of latent representations is set to 32.
The values of $\alpha_1$ and $\alpha_2$ in Eq.~\eqref{eq:filter_client} are set to $25$ and $75$, and the hyper-parameters $\kappa_1$, $\kappa_2$, $\kappa_3$, and $\kappa_4$ are set to values of 0.1, 0.1, 0.3, and 0.5 respectively.
Prior to executing graph clustering, we use HDBSCAN~\cite{campello2013density} to cluster client features $\mathcal{X}$, thereby determining the number of clusters $Q$.

\noindent \textbf{Datasets}.
Consistent with prior researches~\cite{bagdasaryan2020how,wang2020attack}, We evaluate G$^2$uardFL on four different tasks:
1) Digit classification on MNIST~\cite{lecun1998gradient} with LeNet,
2) Image classification on CIFAR-10~\cite{krizhevsky2009learning} with VGG-9~\cite{simonyan2015very},
3) Sentiment classification on Sentiment-140~\cite{sahni2017efficient} with LSTM~\cite{hochreiter1997long},
and 4) Next word prediction on Reddit~\cite{mcmahan2017communication} with LSTM.
We distribute the training data uniformly to construct the local data $\mathcal{D}_{c_i}$ for benign client $c_i$, unless otherwise specified.
All other hyper-parameters are specified in Appendix~\ref{subsec:detail_dataset}.

For the poisoned data $\mathcal{D}^{\mathcal{A}}$, we apply the following strategies:
In the first task, we choose images of \emph{digit 7} from Ardis~\cite{kusetogullari2020ardis} and label them as \emph{1}.
In the second task, we gather images of \emph{planes of Southwest Airline} and label them as \emph{truck}.
In the third task, we collect tweets containing the name of Greek film director \emph{Yorgos Lanthimos} along with positive sentiment comments and label them \emph{negative}.
Finally, in the fourth task, we replace the suffix of sentences with \emph{pasta from Astoria is} and set the last word as \emph{delicious}.

\noindent \textbf{Baselines}.
For comparison, we utilize state-of-the-art defenses, including FLAME~\cite{nguyen2022flame}, DeepSight~\cite{rieger2022deep}, FoolsGold~\cite{fung2020limitations}, (Multi-)Krum~\cite{blanchard2017machine}, RLR~\cite{ozdayi2021defending}, FLTrust~\cite{cao2021fltrust}, NDC~\cite{sun2019can}, RFA~\cite{pillutla2019robust}, and Weak DP~\cite{dwork2014algorithmic} are all employed for comparison.
All these methods are discussed in Section~\ref{sec:related_work}, and additional details can be found in Appendix~\ref{subsec:detail_baseline}.

\noindent \textbf{Backdoor Attacks}.
We select five typical backdoor attacks to evaluate the robustness of G$^2$uardFL:
\begin{itemize}[leftmargin=10pt]
    \item Black-box Attack~\cite{wang2020attack}, a data poisoning attack where the attacker does not have extra access to the FL process and trains using $\mathcal{D}^{\mathcal{A}}$ like other benign clients. 
    \item PGD \emph{with/without} replacement~\cite{wang2020attack}, a model poisoning attack where the attacker uses PGD to limit the divergence between poisoned models and the global model, with/without modifying weights of poisoned models to eliminate contributions of other benign clients.
    \item Constrain-and-scale~\cite{bagdasaryan2020how}, a model poisoning attack that alters the objective function to evade anomaly detection methods such as adding the limitation of $L_p$ distances. 
    \item DBA~\cite{xie2020dba}, a data poisoning attack that decomposes the trigger pattern into sub-patterns and distributes them among several malicious clients for implantation.
    \item 3DFed~\cite{li20233dfed} employs the attack feedback in the previous round to obtain the higher attack performance by implanting indicators into a backdoor model.
\end{itemize}
Detailed settings are outlined in Appendix~\ref{subsec:detail_backdoor_attack}.

Unless otherwise mentioned, our simulated FL system consists of 200 clients, with the orchestration server randomly selecting 10 clients per round.
The selected clients per round train their local models for 2 local gradient descent steps.
50 out of 200 clients are compromised by the attacker, i.e., $\bs{PMR = 0.25}$, and malicious clients have an equal proportion of normal data and backdoored data, i.e., $PDR = 0.5$, unless otherwise stated.
Malicious clients conduct backdoor attacks whenever they are chosen by the orchestration server in a given round.
Following prior studies~\cite{bagdasaryan2020how,xie2020dba,zhang2022flip}, all attacks occur after the global model converges.
In all evaluations, we set the maximum FL round to 600.

\noindent \textbf{Evaluation Metrics}.
We use several metrics to assess defense effectiveness, including Attack Success Rate (ASR), primary task accuracy (ACC), and Defense Score (DS).
ASR measures the proportion of backdoored samples misclassified as the target label.
ACC represents the percentage of correct predictions on benign samples.
As our goal is to decline ASR while upholding ACC, $DS = \nicefrac{2 ACC (1 - ASR)}{ACC + (1 - ASR)}$ computed similarly to F1 Score~\cite{sokolova2009systematic} is introduced as a combined metric to balance ASR and ACC.
Additionally, we evaluate malicious client detection using Precision, Recall, and F1 Score.

\begin{table*}[t]
    \centering
    \caption{Robustness of G$^2$uardFL for various backdoor attacks(\%).}\label{tab:robust_result}
    \renewcommand\arraystretch{1.5}
    \scalebox{0.62}{
    \begin{threeparttable}
        \begin{tabular}{p{1.8cm}p{3.0cm}<{\centering}p{2.1cm}<{\centering} p{2.0cm}<{\centering} p{2.0cm}<{\centering} p{2.0cm}<{\centering} p{2.0cm}<{\centering} p{2.0cm}<{\centering} p{2.0cm}<{\centering} p{2.0cm}<{\centering} p{2.0cm}<{\centering}}
            \toprule[1.2pt]
            \multicolumn{3}{c}{\multirow{2}{*}{Attacks}} & \multicolumn{2}{c}{MNIST} & \multicolumn{2}{c}{CIFAR-10} & \multicolumn{2}{c}{Sentiment-140} & \multicolumn{2}{c}{Reddit} \\
            ~ & ~ & ~ & ASR & ACC & ASR & ACC & ASR & ACC & ASR & ACC \\
            \midrule[1pt]
            \multicolumn{1}{c}{\multirow{10}{*}{\parbox{1.8cm}{\centering Typical Attacks}}} & \multicolumn{1}{c}{\multirow{2}{*}{\parbox{3.0cm}{\centering Black-box \\ Attack~\cite{wang2020attack}}}} & \emph{no-defense} & 34.00 & 99.30 & 17.22 & 85.38 & 99.01 & 65.69 & 100.00 & 22.02 \\
            \multicolumn{1}{c}{~} & \multicolumn{1}{c}{~} & G$^2$uardFL & 00.00 {\scriptsize $\downarrow 34.00$} & 99.31 & 01.11 {\scriptsize $\downarrow 16.11$} & 84.05 & 07.92 {\scriptsize $\downarrow 91.09$} & 74.86 & 00.00 {\scriptsize $\downarrow 100.00$} & 22.63 \\
            \multicolumn{1}{c}{~} & \multicolumn{1}{c}{\multirow{2}{*}{\parbox{3.0cm}{\centering PGD \emph{w/o} \\ replacement~\cite{wang2020attack}}}} & \emph{no-defense} & 41.00 & 99.28 & 22.78 & 85.22 & 99.01 & 65.54 & 100.00 & 22.66 \\
            \multicolumn{1}{c}{~} & \multicolumn{1}{c}{~} & G$^2$uardFL & 00.00 {\scriptsize $\downarrow 41.00$} & 99.35 & 01.67 {\scriptsize $\downarrow 21.11$} & 83.97 & 07.92 {\scriptsize $\downarrow 91.09$} & 74.93 & 00.00 {\scriptsize $\downarrow 100.00$} & 22.75 \\
            \multicolumn{1}{c}{~} & \multicolumn{1}{c}{\multirow{2}{*}{\parbox{3.0cm}{\centering PGD \emph{w/} \\ replacement~\cite{wang2020attack}}}} & \emph{no-defense} & 68.00 & 99.26 & 47.78 & 84.97 & 100.00 & 64.51 & 100.00 & 22.58 \\
            \multicolumn{1}{c}{~} & \multicolumn{1}{c}{~} & G$^2$uardFL & 00.00 {\scriptsize $\downarrow 68.00$} & 99.33 & 01.67 {\scriptsize $\downarrow 46.11$} & 84.84 & 07.92 {\scriptsize $\downarrow 92.08$} & 74.12 & 00.00 {\scriptsize $\downarrow 100.00$} & 22.48 \\
            \multicolumn{1}{c}{~} & \multicolumn{1}{c}{\multirow{2}{*}{\parbox{3.0cm}{\centering Constrain-and-scale~\cite{bagdasaryan2020how}}}} & \emph{no-defense} & 67.00 & 99.19 & 29.44 & 85.20 & 100.00 & 65.22 & 100.00 & 22.18 \\
            \multicolumn{1}{c}{~} & \multicolumn{1}{c}{~} & G$^2$uardFL & 00.00 {\scriptsize $\downarrow 67.00$} & 99.29 & 01.67 {\scriptsize $\downarrow 68.00$} & 84.44 & 08.91 {\scriptsize $\downarrow 91.09$} & 74.83 & 00.00 {\scriptsize $\downarrow 100.00$} & 22.86 \\
            \multicolumn{1}{c}{~} & \multicolumn{1}{c}{\multirow{2}{*}{\parbox{3.0cm}{\centering DBA~\cite{xie2020dba}}}} & \emph{no-defense} & 100.00 & 99.16 & 99.14 & 82.03 & N/A & N/A & N/A & N/A \\
            \multicolumn{1}{c}{~} & \multicolumn{1}{c}{~} & G$^2$uardFL & 18.63 {\scriptsize $\downarrow 91.37$} & 99.28 & 18.14 {\scriptsize $\downarrow 81.00$} & 84.21 & N/A & N/A & N/A & N/A \\
            \multicolumn{1}{c}{~} & \multicolumn{1}{c}{\multirow{2}{*}{\parbox{3.0cm}{\centering 3DFed~\cite{li20233dfed}}}} & \emph{no-defense} & 98.18 & 97.52 & 96.16 & 73.08 & N/A & N/A & N/A & N/A \\
            \multicolumn{1}{c}{~} & \multicolumn{1}{c}{~} & G$^2$uardFL & 09.98 {\scriptsize $\downarrow 88.20$} & 97.84 & 10.61 {\scriptsize $\downarrow 85.55$} & 73.05 & N/A & N/A & N/A & N/A \\
            \midrule[1pt]
            \multicolumn{1}{c}{\multirow{6}{*}{\parbox{1.8cm}{\centering Adaptive Attacks}}} & \multicolumn{1}{c}{\multirow{2}{*}{\parbox{3.0cm}{\centering Dynamic PDR}}} & \emph{no-defense} & 85.00 & 99.31 & 58.33 & 85.13 & 100.00 & 64.59 & 100.00 & 22.69 \\
            \multicolumn{1}{c}{~} & \multicolumn{1}{c}{~} & G$^2$uardFL & 00.00 {\scriptsize $\downarrow 85.00$} & 99.36 & 03.89 {\scriptsize $\downarrow 54.44$} & 83.78 & 07.92 {\scriptsize $\downarrow 92.08$} & 74.38 & 00.00 {\scriptsize $\downarrow 100.00$} & 22.68 \\
            \multicolumn{1}{c}{~} & \multicolumn{1}{c}{\multirow{2}{*}{\parbox{3.0cm}{\centering Model Obfuscation}}} & \emph{no-defense} & 100.00 & 98.41 & 100.00 & 38.85 & 100.00 & 64.57 & 100.00 & 13.83 \\
            \multicolumn{1}{c}{~} & \multicolumn{1}{c}{~} & G$^2$uardFL & 00.00 {\scriptsize $\downarrow 100.00$} & 99.30 & 03.89 {\scriptsize $\downarrow 96.11$} & 83.93 & 10.89 {\scriptsize $\downarrow 89.11$} & 75.63 & 00.00 {\scriptsize $\downarrow 100.00$} & 21.97 \\
            \multicolumn{1}{c}{~} & \multicolumn{1}{c}{\multirow{2}{*}{\parbox{3.0cm}{\centering Randomized Attack}}} & \emph{no-defense} & 62.00 & 99.20 & 24.44 & 85.56 & 100.00 & 64.82 & 100.00 & 22.53 \\
            \multicolumn{1}{c}{~} & \multicolumn{1}{c}{~} & G$^2$uardFL & 00.00 {\scriptsize $\downarrow 62.00$} & 99.28 & 01.11 {\scriptsize $\downarrow 23.33$} & 84.44 & 06.93 {\scriptsize $\downarrow 93.07$} & 74.94 & 00.00 {\scriptsize $\downarrow 100.00$} & 22.54 \\
            \bottomrule[1.2pt]
        \end{tabular}
    \end{threeparttable}
    }
    \vspace{-10pt}
\end{table*}

\subsection{Evaluating Effectiveness of G\texorpdfstring{$^2$}uardFL}\label{subsec:evaluate_effectiveness}

In this subsection, we evaluate the performance of G$^2$uardFL in comparison to nine state-of-the-art defense mechanisms across four distinct tasks.
Considering the trade-off between attack effectiveness and primary task performance, as discussed in Subsection~\ref{subsec:evaluate_robustness}, we choose the \emph{3DFed} attack~\cite{li20233dfed} on the MNIST and CIFAR-10 datasets, while selecting the \emph{PGD with replacement} attack~\cite{wang2020attack} for the Sentiment-140 and Reddit datasets.
Furthermore, we initially assume that client data is independently and identically distributed to assess the peak defensive performance.
Subsequently, we demonstrate the performance of G$^2$uardFL under varying degrees of data heterogeneity in Subsection~\ref{subsec:evaluate_robustness}.
The experimental results are presented in Table~\ref{tab:effective_result}, wherein \emph{Benign setting} signifies the absence of backdoor attacks, and \emph{No defense} implies the presence of malicious clients without any defensive countermeasures.
Our observations are summarized as follows:

1) G$^2$uardFL significantly mitigates the effectiveness of attacks while concurrently maintaining a commendably high ACC, thereby achieving the highest DS across four tasks.
This can be attributed to G$^2$uardFL's capacity to construct an attributed graph that captures inter-client relationships and employs graph-based clustering algorithms to aggregate normal weights while excluding poisoned ones.

2) Deployed defenses can paradoxically result in higher ASRs compared to scenarios without defenses.
Since defenses classify a subset of clients as malicious ones, thereby rejecting these local models (e.g., Krum) or reducing weights (e.g., FoolsGold) during aggregation, aggregated models would be poisoned if defense mechanisms erroneously identify malicious clients.

3) Tasks related to textual data are more susceptible to attacks and pose a more formidable challenge to mitigate backdoor attacks, aligning with the findings of Bagdasaryan \emph{et al}.~\cite{bagdasaryan2020how}.
This susceptibility is because of the greater dissimilarity between backdoored samples of textual data and benign clients' data distribution in comparison to image data.
Sentences within backdoored samples in Sentiment-140 and Reddit are less likely to appear in benign clients' local data, rendering backdoors resistant to being overwritten by model updates from other benign clients.

\begin{table}[t]
    \centering
    \caption{Results of malicious client detection(\%).}\label{tab:detection_malicious}
    \renewcommand\arraystretch{1.5}
    \scalebox{0.60}{
    \begin{threeparttable}
        \begin{tabular}{p{2.0cm} p{1.4cm}<{\centering} p{1.4cm}<{\centering} p{1.4cm}<{\centering} p{1.4cm}<{\centering} p{1.4cm}<{\centering} p{1.4cm}<{\centering}}
            \toprule[1.2pt]
            \multirow{2}{*}{\parbox{2.0cm}{\centering Defense}} & \multicolumn{3}{c}{MNIST} & \multicolumn{3}{c}{CIFAR-10} \\
            ~ & Precision & Recall & F1 Score & Precision & Recall & F1 Score \\
            \midrule[1pt]
            FLAME~\cite{nguyen2022flame} & 52.40 & 86.79 & 65.34 & 31.96 & 53.20 & 39.93\\
            DeepSight~\cite{rieger2022deep} & 97.93 & 97.46 & 97.70 & 85.12 & 90.13 & 87.56 \\
            G$^2$uardFL (\textbf{ours}) & 78.49 & 82.84 & 80.61 & 93.26 & 67.62 & 78.40 \\
            G$^2$uardFL${}^\prime$ & 87.92 & 93.66 & 90.70 & 92.22 & 97.55 & 94.81 \\
            \bottomrule[1.2pt]
            \multirow{2}{*}{\parbox{2.0cm}{\centering Defense}} & \multicolumn{3}{c}{Sentiment-140} & \multicolumn{3}{c}{Reddit} \\
            ~ & Precision & Recall & F1 Score & Precision & Recall & F1 Score \\
            \midrule[1pt]
            FLAME~\cite{nguyen2022flame} & 60.04 & 83.24 & 69.76 & 63.72 & 82.20 & 71.79 \\
            DeepSight~\cite{rieger2022deep} & 00.00 & 00.00 & 00.00 & N/A & N/A & N/A \\
            G$^2$uardFL (\textbf{ours}) & 94.77 & 95.16 & 94.97 & 89.92 & 89.04 & 89.48 \\
            G$^2$uardFL${}^\prime$ & 93.79 & 96.70 & 95.22 & 92.95 & 97.39 & 95.12 \\
            \bottomrule[1.2pt]
        \end{tabular}
    \end{threeparttable}
    }
    \vspace{-10pt}
\end{table}

Additionally, we evaluate the performance of G$^2$uardFL in detecting malicious clients.
The results, comprehensively presented in Table~\ref{tab:detection_malicious}, reveal the following key insights:

1) G$^2$uardFL consistently demonstrates accurate detection of malicious clients across all datasets, consistently achieving F1 Scores exceeding 78\%.
This underscores the potency of low-dimensional latent representations learned by graph-based clustering algorithms, which exhibit heightened representational power relative to high-dimensional model weights adopted by FLAME and DeepSight.

2) Although FLAME exhibits a commendable performance in identifying malicious clients, its effectiveness in defending against backdoor attacks is relatively lower.
This can be attributed to the dynamic nature of client selection in each FL round, where clients are randomly chosen by the orchestration server in this paper.
Furthermore, the number of chosen malicious clients varies over rounds.
However, FLAME employs a straightforward strategy of designating clusters with smaller sizes as malicious and filtering out malicious ones during the aggregation.
Consequently, there exists a notable probability that malicious clients dominate in a particular round, leading FLAME to erroneously accept these malicious clusters to construct the global model.

3) In the MNIST and CIFAR-10 datasets, G$^2$uardFL, while potentially yielding a slightly lower F1 Score than DeepSight, outperforms in terms of defense effectiveness.
This divergence can be attributed to the adaptive poison eliminating module, which removes previously implanted backdoors, improves the difficulty of implanting backdoors in subsequent rounds, and finally enhances the similarity between normal and poisoned models.
To validate this conjecture, we evaluate G$^2$uardFL without the adaptive poison eliminating module, denoted as $\text{G$^2$uardFL}^\prime$.
The outcomes affirm our hypothesis, as the graph-based detection method utilizes clustering algorithms on attributed graphs to acquire more representative latent representations.

\subsection{Evaluating Robustness of G\texorpdfstring{$^2$}uardFL}\label{subsec:evaluate_robustness}

In this subsection, we investigate the robustness of G$^2$uardFL against six standard backdoor attacks, three adaptive attacks, and diverse degrees of non-iid data.

\noindent \textbf{Typical Attacks}.
We assess G$^2$uardFL's resilience against state-of-the-art backdoor attacks including Black-box Attack, PGD without replacement, PGD with replacement, Constrain-and-scale, DBA, and 3DFed, across multiple datasets.
Note that, given that 3DFed designs indicators to acquire attack feedback, we use the model specified in \cite{li20233dfed} as the global model in our evaluation.
Hence, the primary task performance of 3DFed differs from that of other attacks.
The results, summarized  in Table~\ref{tab:robust_result}, reveal the insights:

1) G$^2$uardFL adeptly thwarts typical attacks while maintaining primary task performance. 
It achieves this by taking into account backdoor attacks that constrain the $L_p$-norm of differences between poisoned models and the global model when forming inter-client relationships, thus effectively countering Constrain-and-scale attacks.

2) DBA, which decomposes triggers in backdoored samples into some local triggers to maximize attack performance, presents a more formidable challenge for G$^2$uardFL compared to the other four attacks.
This aligns with findings from Nguyen \emph{et al}.~\cite{nguyen2023backdoor}.
However, DBA is only designed for image-related tasks, limited by its trigger design.

3) PGD with replacement achieves a favorable balance between achieving a high ASR and maintaining primary task performance.
Consequently, we designate it as the default backdoor attack in our evaluations.

\begin{figure}[t]
    \centering
    \includegraphics[width=0.90\linewidth]{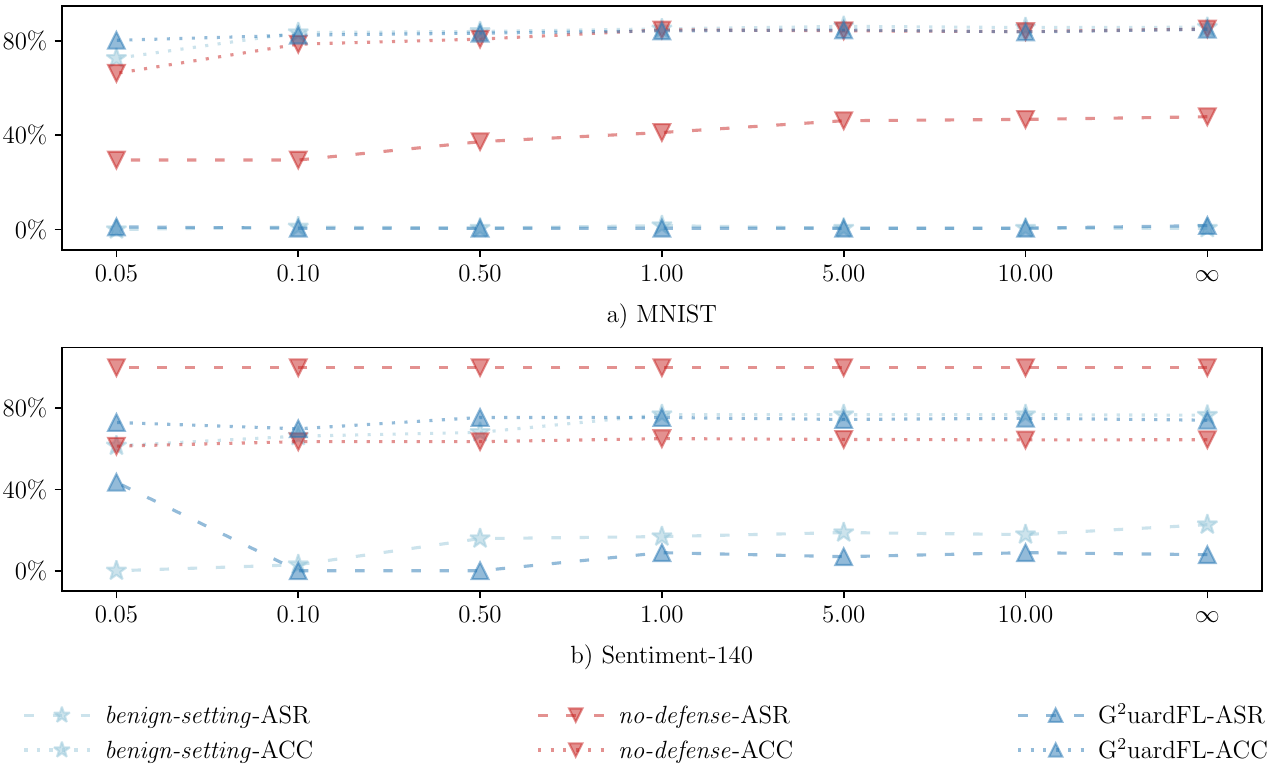}
    \caption{The impact of data heterogeneity in G$^2$uardFL on CIFAR-10 and Sentiment-140.}\label{fig:cifar_and_sentiment_data}
    \vspace{-10pt}
\end{figure}

\noindent \textbf{Adaptive Attacks}.
We assess G$^2$uardFL's resilience against adaptive attacks, where attackers possess sufficient knowledge of G$^2$uardFL and aim to circumvent its defenses.
We examine several attack strategies and scenarios:
\begin{itemize}[leftmargin=10pt]
    \item \noindent \emph{Dynamic PDR}.
    Attackers can manipulate models to minimize differences between normal and poisoned models, leading to their inclusion in the same cluster during graph clustering.
    This is achieved by assigning a random PDR ranging from 5\% to 20\% to each malicious client.
    \item \noindent \emph{Model Obfuscation}.
    Attackers can introduce noise to poisoned models before transmission to the orchestration server, rendering them more challenging to detect.
    We set the noise level following previous work~\cite{nguyen2022flame} to 0.034 to avoid significantly impacting ASR.
    \item \noindent \emph{Randomized Attack}.
    G$^2$uardFL relies on both cluster sizes and benign scores to identify malicious clusters.
    In this scenario, malicious clients selected in an FL round have a 50\% probability of launching attacks, undermining the utility of historical detection results.
\end{itemize}
The results in Table~\ref{tab:robust_result} yield the following observations:

1) G$^2$uardFL achieves low ASR and high ACC in all cases (e.g., Model Obfuscation in CIFAR-10).
This success can be attributed to the graph construction module, which effectively captures inter-client relationships, and the dynamic filtering and clipping module, which accurately discerns benign and malicious clusters.
The removal of poisoned models during the aggregation process ensures that they do not adversely impact the performance of the aggregated model in the primary task.

2) The Randomized Attack has a limited impact on G$^2$uardFL's performance due to the introduction of Eq.~\eqref{eq:smooth_graph}.
This equation serves to smooth the constructed attributed graph, thereby mitigating the instability of detection results.

\begin{table}[t]
    \centering
    \caption{The impact of PMR on G$^2$uardFL(\%).}\label{tab:influence_pmr}
    \scalebox{0.55}{
    \renewcommand\arraystretch{1.5}
    \begin{tabular}{p{1.4cm}<{\centering} p{1.4cm}<{\centering} p{1.4cm}<{\centering} p{1.4cm}<{\centering} p{1.4cm}<{\centering} p{1.4cm}<{\centering} p{1.4cm}<{\centering} p{1.4cm}<{\centering}}
        \toprule[1.2pt]
        \multicolumn{2}{c}{PMR} & 0.00 & 0.05 & 0.10 & 0.15 & 0.20 & 0.25 \\
        \midrule[1pt]
        \multicolumn{1}{c}{\multirow{2}{*}{\parbox{1.3cm}{\centering \emph{no-defense}}}} & ASR & 00.56 & 37.22 & 40.00 & 41.11 & 43.89 & 47.78  \\
        ~ & ACC & 85.72 & 82.22 & 85.66 & 85.68 & 85.15 & 84.97 \\
        \multicolumn{1}{c}{\multirow{2}{*}{\parbox{1.3cm}{\centering G$^2$uardFL}}} & ASR & 02.78 & 00.56 & 00.00 & 00.56 & 00.56 & 01.67 \\
        ~ & ACC & 85.05 & 84.91 & 84.48 & 84.46 & 84.84 & 84.04 \\
        \midrule[1.2pt]
        \multicolumn{2}{c}{PMR} & 0.30 & 0.35 & 0.40 & 0.45 & 0.50 & 0.55 \\
        \midrule[1pt]
        \multicolumn{1}{c}{\multirow{2}{*}{\parbox{1.3cm}{\centering \emph{no-defense}}}} & ASR & 48.33 & 47.22 & 48.89 & 55.56 & 62.22  & 65.00  \\
        ~ & ACC & 84.24 & 85.11 & 84.33 & 82.49 & 83.28 & 83.00 \\
        \multicolumn{1}{c}{\multirow{2}{*}{\parbox{1.3cm}{\centering G$^2$uardFL}}} & ASR & 00.00 & 03.89 & 17.22 & 13.89 & 16.11  & 78.33  \\
        ~ & ACC & 84.62 & 83.88 & 83.32 & 83.18 & 83.40 & 82.30 \\
        \bottomrule[1.2pt]
    \end{tabular}
    }
    \vspace{-2pt}
\end{table}

\begin{table}[t]
    \centering
    \caption{The impact of $M$ on G$^2$uardFL(\%).}\label{tab:influence_m}
    \scalebox{0.55}{
    \renewcommand\arraystretch{1.5}
    \begin{tabular}{p{1.4cm}<{\centering} p{1.4cm}<{\centering} p{1.8cm}<{\centering} p{1.8cm}<{\centering} p{1.8cm}<{\centering} p{1.8cm}<{\centering} p{1.8cm}<{\centering}}
        \toprule[1.2pt]
        \multirow{2}{*}{\parbox{1.3cm}{\centering M}} & \multicolumn{2}{c}{\emph{Benign setting}} & \multicolumn{2}{c}{\emph{no-defense}} & \multicolumn{2}{c}{G$^2$uardFL} \\
        ~ & ASR & ACC & ASR & ACC & ASR & ACC  \\
        \midrule[1pt]
        5\% & 00.56 & 85.89 & 48.33 & 83.80 & 01.67 & 84.16 \\
        10\% & 00.56 & 85.89 & 47.78 & 83.48 & 01.11 & 84.17 \\
        15\% & 00.56 & 86.08 & 50.56 & 83.62 & 01.11 & 83.78 \\
        \bottomrule[1.2pt]
    \end{tabular}
    }
    \vspace{-10pt}
\end{table}

\noindent \textbf{Data Heterogeneity}.
The variability in clients' data distribution can potentially influence G$^2$uardFL's performance, as the graph construction module evaluates the disparity between normal and poisoned model weights.
To simulate data heterogeneity, we introduce data partitioning~\cite{hsu2019measuring} by sampling diverse proportions of data to clients.
Specifically, we use a Dirichlet distribution parameterized by $\alpha$ to allocate data to clients, ranging from highly non-iid ($\alpha \rightarrow 0$) to uniformly distributed ($\alpha \rightarrow \infty$).
We choose $\alpha \in \{ 0.05, 0.10, 0.50, 1.00, 5.00, 10.00 \}$ and the evaluation results in Fig.~\ref{fig:cifar_and_sentiment_data} with the \emph{PGD with replacement} attack show:

1) G$^2$uardFL effectively defends against backdoor attacks, with a marginal drop in primary task performance as data heterogeneity increases.
The attributed graph construction ensures that differences between benign and malicious clients outweigh differences among benign clients, facilitating the filtration of poisoned weights regardless of data heterogeneity levels.

2) At $\alpha = 0.05$, the ASR of Sentiment-140 experiences a significant decline under G$^2$uardFL's defense.
This phenomenon is attributed to the fact that numerous benign clients possess only one sample at $\alpha = 0.05$, as evidenced in Fig.~\ref{fig:vis_sentiment}.
If G$^2$uardFL erroneously accepts poisoned weights in rounds where they prevail, it becomes challenging to eliminate the backdoors from the aggregated global model due to the limited number of samples on benign clients.

\begin{table}[t]
    \centering
    \caption{Impacts of $\kappa_1$, $\kappa_2$, $\kappa_3$ and $\kappa_4$ on G$^2$uardFL(\%).}\label{tab:influence_factor}
    \renewcommand\arraystretch{1.5}
    \scalebox{0.55}{
    \begin{tabular}{p{1.4cm}<{\centering} p{1.4cm}<{\centering} p{1.4cm}<{\centering} p{1.4cm}<{\centering} p{1.4cm}<{\centering} p{1.4cm}<{\centering} p{1.4cm}<{\centering} p{1.4cm}<{\centering}}
        \toprule[1.2pt]
        $\kappa_1$ & 0.0 & 0.1 & 0.2 & 0.3 & 0.4 & 0.5 & 0.6 \\
        \midrule[1pt]
        ASR & 01.67 & 01.67 & 00.56 & 01.67  & 00.56  & 00.56  & 00.00 \\
        ACC & 84.01 & 84.84 & 83.46 & 83.96 & 83.60 & 84.20 & 84.33 \\
        \midrule[1.2pt]
        $\kappa_2$ & 0.0 & 0.1 & 0.2 & 0.3 & 0.4 & 0.5 & 0.6   \\
        \midrule[1pt]
        ASR  & 00.00  & 01.67  & 00.00  & 01.11  & 01.11  & 01.11  & 01.11  \\
        ACC  & 83.89 & 84.84 & 84.05 & 83.47 & 84.28 & 84.33 & 84.65 \\
        \midrule[1.2pt]
        $\kappa_3$ & 0.1 & 0.2 & 0.3 & 0.4 & 0.5 & 0.6 & 0.7   \\
        \midrule[1pt]
        ASR  & 01.11  & 03.89  & 01.67  & 03.89  & 03.89  & 05.00  & 02.78  \\
        ACC  & 84.06 & 84.11 & 84.84 & 83.95 & 84.11 & 84.07 & 84.71 \\
        \midrule[1.2pt]
        $\kappa_4$ & 0.1 & 0.2 & 0.3 & 0.4 & 0.5 & 0.6 & 0.7   \\
        \midrule[1pt]
        ASR  & 00.00  & 05.00  & 01.11  & 01.11  & 01.67  & 00.00  & 00.00  \\
        ACC  & 84.45 & 84.28 & 84.26 & 84.14 & 84.84 & 83.89 & 83.95 \\
        \bottomrule[1.2pt]
    \end{tabular}
    }
    \vspace{-10pt}
\end{table}

\subsection{Evaluating Sensitivity of G\texorpdfstring{$^2$}uardFL}\label{subsec:evaluate_sensitivity}

\noindent \textbf{PMR}.
We assess G$^2$uardFL's performance under different PMR values on the CIFAR-10 dataset.
The results in Table~\ref{tab:influence_pmr} demonstrate that G$^2$uardFL effectively defends against backdoor attacks as long as the number of malicious clients in the FL system remains below 50\% (PMR less than 0.5).
This robustness arises from the dynamic filtering and clipping module's ability to utilize benign scores for identifying malicious clients.
However, when PMR exceeds 0.5, and malicious clients constitute the majority in the FL system, G$^2$uardFL's defense capabilities diminish.

\noindent \textbf{Number of Clients} $M$.
In practical FL scenarios, clients may join or leave the system dynamically during the FL process.
In our evaluation, we simulate scenarios where a certain proportion of clients go offline or join the FL system (i.e., 5\%, 10\%, and 15\%).
The results, as presented in Table~\ref{tab:influence_m}, indicate that the number of clients has a negligible impact on G$^2$uardFL's performance.

\noindent $\bs{\kappa_1}$, $\bs{\kappa_2}$, $\bs{\kappa_3}$ \textbf{and} $\bs{\kappa_4}$.
We explore the influence of hyper-parameters on G$^2$uardFL's performance, as detailed in Table~\ref{tab:influence_factor}.
The results demonstrate that the defense results remain stable even when hyper-parameters vary.
Among these hyper-parameters, $\kappa_3$ exerts a more significant impact on G$^2$uardFL's performance.
This is primarily because cluster sizes change discretely, unlike benign scores, necessitating a delicate balance between these two factors to accurately identify malicious clients.

\section{Discussion}

\noindent \textbf{Advantages of G$^2$uardFL over other defenses (e.g., FLAME~\cite{nguyen2022flame} and DeepSight~\cite{rieger2022deep}).}
A key factor lies in benign scores.
This paper considers the practical complexity setting where the server dynamically selects clients for participation in each round.
The variability in the number of malicious clients selected during a round challenges traditional defenses like FLAME and DeepSight.
To address this, G$^2$uardFL introduces benign scores, leveraging historical detection results to identify malicious client clusters accurately.
The motivation of this strategy is that, while poisoned models may dominate in specific rounds, they constitute a minority on average.

\noindent \textbf{Side effects on primary task performance.}
Indeed, G$^2$uardFL introduces a slight reduction in primary task performance when initiating G$^2$uardFL at the beginning of FL training, as evident in Table~\ref{tab:convergence_analysis}.
However, this effect can be mitigated strategically.
The adaptive poison eliminating module empowers G$^2$uardFL to neutralize the influence of previously embedded backdoors effectively.
Defenders can prudently initiate G$^2$uardFL when the global model has already converged, minimizing potential side effects on primary task performance.

\noindent \textbf{Practicalness of G$^2$uardFL}.
G$^2$uardFL eliminates constraints on aggregation strategies that require all clients to participate in the aggregation and obviates the need for the server to validate client-transmitted weights using collected data.
In all experiments, the graph clustering utilizes latent embeddings of only 32 dimensions, delivering exceptional detection results.
This suggests that G$^2$uardFL can function on servers without demanding significant computational resources, which shows its potential practicalness.

\section{Related Work}\label{sec:related_work}

Existing defenses can broadly be divided into two categories: anomaly detection-based approaches and anomaly detection-free approaches.
The first group identifies poisoned weights as anomalous data in the distribution of client weights and excludes them from aggregation.
In contrast, the second group employs various methods to reduce the influence of poisoned weights during the aggregation process.

\noindent \textbf{Anomaly Detection-based Approaches}.
These methods generally assume that weights from malicious clients are similar and utilize anomaly detection techniques to identify poisoned weights~\cite{blanchard2017machine,fung2020limitations,nguyen2022flame,rieger2022deep}.
Krum and Multi-Krum~\cite{blanchard2017machine} aggregate client models by choosing a local model with the smallest average Euclidean distance to a portion of other models as the new global model.
FoolsGold~\cite{fung2020limitations} assumes that poisoned models always exhibit higher similarity and employs the cosine similarity function to measure the similarity between model weights.
Other defenses such as DeepSight~\cite{rieger2022deep} and FLAME~\cite{nguyen2022flame} assess differences in structure, weights, and outputs of local models and use the HDBSCAN algorithm to detect poisoned models.

However, Krum, Multi-Krum, and FoolsGold only operate under specific assumptions about the underlying data distributions.
Krum and Multi-Krum assume that data from benign clients are independent and identically distributed (iid), while FoolsGold assumes that backdoored samples are iid.
Furthermore, the weight distance measurement used by DeepSight and FLAME cannot provide a comprehensive representation of client weights.
As a result, as demonstrated in Subsection~\ref{subsec:evaluate_effectiveness}, these approaches may either reject normal weights, causing degradation in global model performance, or accept poisoned weights, leaving backdoors effective.

\noindent \textbf{Anomaly Detection-free Approaches}.
These methods employ strategies such as robust aggregation rules~\cite{ozdayi2021defending,pillutla2019robust,sun2019can}, differential privacy~\cite{dwork2014algorithmic} and local validation~\cite{cao2021fltrust}, to reduce the impact of backdoors.
Norm Different Clipping (NDC)~\cite{sun2019can} examines the norm difference between the global model and client model weights, and clips the norm of model updates that exceed a threshold.
RFA~\cite{pillutla2019robust} computes a weighted geometric median using the smoothed Weiszfeld's algorithm to aggregate local models.
RLR~\cite{ozdayi2021defending} adjusts the learning rate of the orchestration server in each round, according to the sign information of client model updates.
Weak DP~\cite{dwork2014algorithmic} applies the Differential Privacy (DP) technique to degrade the performance of local models from malicious clients, where Gaussian noise with a small standard deviation is added to the model aggregated by standard mechanisms, such as FedAvg.
FLTrust~\cite{cao2021fltrust} maintains a model in the orchestration server and assigns a lower trust score to the client whose update direction deviates more from the direction of the server model update.

However, anomaly detection-free approaches do not completely discard local models of malicious clients, allowing a substantial portion of poisoned weights to influence the aggregated global model.
Therefore, these methods are only effective when the number of malicious clients is relatively small, such as less than 10\%~\cite{nguyen2023backdoor}.
FLTrust assumes the server collects a small number of clean samples, which violates the essence of FL~\cite{zhang2022flip}.
Additionally, as shown in Subsection~\ref{subsec:evaluate_effectiveness}, DP-based methods result in a degradation of the primary task performance due to the added noise.

\section{Conclusion}\label{sec:conclusion}

In this paper, we have proposed G$^2$uardFL, an effective and robust framework safeguarding FL against backdoor attacks via attributed client graph clustering.
The detection of malicious clients has been reformulated as an attributed graph clustering problem.
In this framework, a graph clustering method has been developed to divide clients into distinct groups based on their similar characteristics and close connections.
An adaptive mechanism has been devised to weaken the effectiveness of backdoors previously implanted in global models.
We have also conducted a theoretical convergence analysis of G$^2$uardFL.
Comprehensive experimental results have demonstrated that G$^2$uardFL can significantly mitigate the impact of backdoor attacks while preserving the performance of the primary task.
Future research will focus on investigating methods to further enhance G$^2$uardFL's effectiveness and defend against more powerful backdoor attacks.

\newpage
{
\bibliographystyle{abbrv}
\bibliography{refs}
}

\appendices

\begin{table}[t]
    \centering
    \caption{List of notations.}\label{tab:notations}
    \renewcommand\arraystretch{1.5}
    \scalebox{0.63}{
    \begin{tabular}{p{0.9cm}<{\centering} p{4.8cm}|p{0.9cm}<{\centering} p{4.8cm}}
        \toprule[1.2pt]
        \textbf{Sign} & \makecell[c]{\textbf{Description}} & \textbf{Sign} & \makecell[c]{\textbf{Description}} \\
        \midrule[1pt]
        $\mathcal{S}$ & Orchestration server & $\mathcal{A}$ & Malicious attacker \\
        $\mathcal{C}_m^r$ & Chosen clients at $r$-th round & $\mathcal{C}_K$ & Malicious clients \\
        $\mathcal{D}_{c_i}$ & Local data of client $c_i$ & $\mathcal{D}^{\mathcal{A}}$ & Backdoored samples \\
        $G^r$ & Global weights at $r$-th round & $\tilde{G}$ & Poisoned global weights \\
        $W_{c_i}^r$ & Local weights of $c_i$ at the $r$-th round & $\Delta W_{c_i}^r$ & Corresponding model updates of $W_i^r$ \\
        $\tilde{W}$ & Poisoned local weights & $F(\cdot)$ & Task model of FL systems \\
        $L$ & The number of $f(\cdot)$ layers & $R$ & The number of iteration rounds \\
        $M$ & The number of all clients & $K$ & The number of malicious clients \\
        $m$ & The number of chosen clients & $y_c$ & Attacker-chosen predictions \\
        \bottomrule[1.2pt]
    \end{tabular}
    }
    \vspace{-10pt}
\end{table}

\section{Proof of Theorem~\ref{thm:converage}}\label{sec:proof}

Our analysis considers each client conducts $E$ stochastic gradient descent updates per FL round.
The main notations used in this paper are summarized in Table~\ref{tab:notations}.

\subsection{Proof of Convergence Rate}

With Assumption~\ref{assump:smooth}, we have
\begin{equation}\label{eq:convergence_smooth}
\rseq{0.80\linewidth}{!}{
\begin{aligned}
\mathbb{E}\left[F\left(G^r\right)\right]-\mathbb{E}\left[F\left(G^{r-1}\right)\right] &\leq \underbrace{\mathbb{E}\left[\left\langle\nabla F\left(G^{r-1}\right), G^r-G^{r-1}\right\rangle\right]}_{H_1} + \\
&\qquad \frac{L}{2} \underbrace{\mathbb{E}\left[\left\|G^r-G^{r-1}\right\|^2\right]}_{H_2}
\end{aligned}
}.
\end{equation}
We need the following three key lemmas which are proved in subsequent subsections.

\begin{lemma}\label{lemma:diff}
Let $\lambda$ be the learning rate of all benign and malicious clients, we have
\begin{equation}\label{eq:diff_lemma_eq}
\rseq{0.65\linewidth}{!}{
\begin{aligned}
& G^r-G^{r-1}=- r_4 \sum_{c_i \in \mathcal{C}^{+}} \sum_{e = 1}^E \nabla F_{c_i}\left(W_{c_i}^{e-1} ; \xi_{c_i}^e\right) + \\
& \qquad r_5 \sum_{c_i \in \mathcal{C}^{-}} \sum_{e = 1}^E \nabla F_{c_i}\left(\tilde{W}_{c_i}^{e-1} ; \tilde{\xi}_{c_i}^e\right)
\end{aligned}
},
\end{equation}
where $r_1=\min (1, \frac{\|G^{+}-G^{r-1}\|_2}{\|G^{-}-G^{r-1}\|_2} )$, $r_2 = \frac{\sum_{c_j \in \mathcal{C}^{-}} \opn{abs}(\bs{s}_{m, j}^r)}{\sum_{c_k \in \mathcal{C}_m^r} \opn{abs}\left(\bs{s}_{m, k}^r\right)}$, $r_3=\gamma r_2 \log (1 + n^r)$, $r_4 = (1 + r_3) \lambda$ and $r_5 = r_1 r_3 \lambda$.
\end{lemma}

\begin{lemma}\label{lemma:inner_product}
Under Assumption~\ref{assump:smooth} to~\ref{assump:norm_bounded}, it holds that
\begin{equation}\label{eq:convergence_inner_product}
\rseq{0.82\linewidth}{!}{
\begin{aligned}
& H_1 \leq \frac{\left(5r_5 - r_4\right)}{2} K E \underbrace{\mathbb{E}\left[\left\|\nabla F\left(G^{r-1}\right)\right\|^2\right]}_{H_3} + r_4 E \sum_{c_i \in \mathcal{C}^{+}} D_i^2 + \\
& 2 r_5 E \sum_{c_i \in \mathcal{C}^{-}} D_i^2+ \frac{ r_5 }{2} K E \zeta^2 + r_4 L^2 \sum_{c_i \in \mathcal{C}^{+}} \sum_{e=1}^E \underbrace{\mathbb{E}\left[\left\|W_{c_i}^{e-1}-G^{r-1}\right\|^2\right]}_{H_4} + \\
& r_5 L^2 \sum_{c_i \in \mathcal{C}^{-}} \sum_{e = 1}^E \underbrace{\mathbb{E}\left[\left\|\tilde{W}_{c_i}^{e-1}-G^{r-1}\right\|^2\right]}_{H_5}
\end{aligned}
}.
\end{equation}
\end{lemma}

\begin{lemma}\label{lemma:diff_norm}
The difference between two global models during two successive rounds is bounded by
\begin{equation}\label{eq:convergence_diff_norm}
\rseq{0.82\linewidth}{!}{
\begin{aligned}
& H_2 \leq 2 r_4^2 K^2 E^2 \zeta^2 + 4 r_4^2 K E L^2 \sum_{c_i \in \mathcal{C}^{+}} \sum_{e = 1}^E H_4 + \\
& 8 r_4^2 K E^2 \sum_{c_i \in \mathcal{C}^{+}} D_i^2 + 8 r_4^2 \sum_{c_i \in \mathcal{C}^{+}} \sum_{e = 1}^E H_3 + 2 r_5^2 K^2 E^2 \zeta^2 + \\
& 4 r_5^2 K E L^2 \sum_{c_i \in \mathcal{C}^{-}} \sum_{e = 1}^E H_5 + 8 r_5^2 K E^2 \sum_{c_i \in \mathcal{C}^{-}} D_i^2 + 8 r_5^2 \sum_{c_i \in \mathcal{C}^{-}} \sum_{e = 1}^E H_3
\end{aligned}
}.
\end{equation}
\end{lemma}

\begin{lemma}\label{lemma:diff_norm_local}
At benign clients, the difference between the local model at each round $r$ and the global model at the previous last round is bounded by
\begin{equation}\label{eq:convergence_diff_norm_local_benign}
\rseq{0.80\linewidth}{!}{
\begin{aligned}
\sum_{e = 1}^{E} H_4 \leq \frac{1}{1 - 2 \lambda^2 E^2 L^2} \left( \lambda^2 E^3 \zeta^2 + 4 \lambda^2 E^3 D_i^2 + 4 \lambda^2 E^3 H_3 \right)
\end{aligned}
},
\end{equation}
where $c_i \in \mathcal{C}^{+}$.
While they are malicious clients, the difference is bounded by
\begin{equation}\label{eq:convergence_diff_norm_local_malicious}
\rseq{0.80\linewidth}{!}{
\begin{aligned}
\sum_{e = 1}^{E} H_5 \leq \frac{1}{1 - 2 \lambda^2 E^2 L^2} \left( \lambda^2 E^3 \zeta^2 + 4 \lambda^2 E^3 D_i^2 + 4 \lambda^2 E^3 H_3 \right)
\end{aligned}
}.
\end{equation}
\end{lemma}

By substituting Eq.~\eqref{eq:convergence_inner_product} into the first term of RHS in Eq.~\eqref{eq:convergence_smooth}, and Eq.~\eqref{eq:convergence_diff_norm} into the second term, and by Eq.~\eqref{eq:convergence_diff_norm_local_benign} and Eq.~\eqref{eq:convergence_diff_norm_local_malicious}, we have
\begin{equation}
\rseq{0.90\linewidth}{!}{
\begin{aligned}
& \left\{\left[\frac{1 - 4 r_3^{\star}}{2} - 2 r_6 (1+2r_3^\star)\right] \lambda K E - 4 \left[{r_3^{\star}}^2 + \left( 1 + r_3^{\star} \right)^2 \right] (1 + r_6) \lambda^2 K^2 E^2 L\right\} H_3 \leq \\
&\mathbb{E}\left[F\left(G^{r-1}\right)\right] - \mathbb{E}\left[F\left(G^r\right)\right] + \\
& \left[\left(1+r_3^{\star}\right) \left(1+2 r_6 \right) \lambda E+4\left(1+r_3^{\star}\right)^2 \left(1+r_6\right) \lambda^2 K E^2 L\right] \sum_{c_i \in \mathcal{C}^{+}} D_i^2 + \\
& \left[2 r_3^{\star} (1 + r_6) \lambda E +4{r_3^{\star}}^2 \left(1+r_6\right) \lambda^2 K E^2 L \right] \sum_{c_i \in \mathcal{C}^{-}} D_i^2+ \\
& \left\{ \frac{r_3^\star + r_6}{2} \lambda K E + r_3^\star r_6 \lambda K E +\lambda^2 K^2 E^2 L (1 + r_6) \left[\left(1+r_3^{\star}\right)^2+ {r_3^\star}^2 \right]\right\} \zeta^2
\end{aligned}
},
\end{equation}
where $r_6 = \nicefrac{2 \lambda^2 E^2 L^2}{1 - 2 \lambda^2 E^2 L^2}$ and $r_3^\star = \max(r_3) = \gamma \log (1 + \max(n^r))$.

Next, summing the above items from $r = 1$ to $R$ and dividing both sides by $\lambda T K$ where $T = RE$ is the total number of local stochastic gradient descent steps yields
\begin{equation}\label{eq:proof_convergence}
\rseq{0.90\linewidth}{!}{
\begin{aligned}
& \underbrace{\left\{\left[\frac{1 - 4 r_3^{\star}}{2} - 2 r_6 (1+2r_3^\star)\right] - 4 \left[{r_3^{\star}}^2 + \left( 1 + r_3^{\star} \right)^2 \right] (1 + r_6) \lambda K E L\right\}}_{H_6} \frac{\sum_{r = 1}^R H_3}{R} \leq \\
& \frac{\mathbb{E}\left[F\left(G^0\right)\right] - \mathbb{E}\left[F\left(G^R \right)\right]}{\lambda T K} + \\
& \underbrace{\left[\frac{\left(1+r_3^{\star}\right) \left(1+2 r_6 \right)}{K}+4\left(1+r_3^{\star}\right)^2 \left(1+r_6\right) \lambda E L\right]}_{H_7} \sum_{c_i \in \mathcal{C}^{+}} D_i^2 + \\
& \underbrace{\left[2 \frac{r_3^{\star}}{K} (1 + r_6) + 4 {r_3^{\star}}^2 \left(1+r_6\right) \lambda E L \right]}_{H_8} \sum_{c_i \in \mathcal{C}^{-}} D_i^2+ \\
& \underbrace{\left\{ \frac{r_3^\star + r_6}{2} + r_3^\star r_6 +\lambda K E L (1 + r_6) \left[\left(1+r_3^{\star}\right)^2+ {r_3^\star}^2 \right]\right\}}_{H_9} \zeta^2
\end{aligned}
},
\end{equation}

Let the learning rate $\lambda=\nicefrac{K^{\frac{1}{2}}}{8 L T^{\frac{1}{2}}}$, the hyper-parameter $\gamma = \nicefrac{\lambda}{\sqrt{K}\log (1 + \max(n^r))}$  and the number of local updating steps $E \leq \nicefrac{T^{\frac{1}{4}}}{K^{\frac{3}{4}}}$, where $T \geq \max \{ 16 K^3, \frac{16}{K}, \frac{16}{L^2} \}$ in order to guarantee $E \geq 1$.
Thus, we have
$$
L T^{\frac{1}{2}} \geq 4, T^{\frac{1}{2}} K^{\frac{1}{2}} \geq 4, L T K^{\frac{1}{2}}=(L T^{\frac{1}{2}})(T^{\frac{1}{2}} K^{\frac{1}{2}}) \geq 16.
$$
Then,
$$
\rseq{0.90\linewidth}{!}{
\begin{aligned}
H_6 &\geq \frac{1}{2}-\frac{1}{16}-\frac{128}{127}\left(\frac{1}{64}+\frac{1}{1024}+\frac{1}{4}+\frac{1}{64}+\frac{1}{1024}\right) \geq \frac{2}{5} \\
H_7 &\leq \frac{129}{127} \times \frac{29}{28 K}+\frac{545}{127 \times 16} \leq \frac{11}{10 K}+\frac{21}{10} \\
H_8 &\leq \frac{8}{127 K}+\frac{1}{127 \times 16} \leq \frac{1}{10 K}+\frac{1}{10} \\
H_9 &\leq \frac{1}{127}+\frac{131}{254} \times \frac{1}{32}+\frac{129}{127} \times \frac{1}{16}+\frac{129}{127} \times \frac{1}{512}+\frac{258}{127} \times \frac{1}{16384} \leq \frac{1}{10}
\end{aligned}
}.
$$

Finally, by substituting $\mathbb{E}[F(G^R)] \geq F(G^\star)$ in Assumption~\ref{assump:smooth} and above coefficients into Eq.~\eqref{eq:proof_convergence}, Theorem~\ref{thm:converage} is proved.

\subsection{Proof of Lemma~\ref{lemma:diff}}

According to Eq.~\eqref{eq:limit_malicious_global} and Eq.~\eqref{eq:aggregate_global}, we have
\begin{equation}\label{eq:diff_proof}
\rseq{0.80\linewidth}{!}{
G^r - G^{r - 1} = (1 + r_3) (G^{+} - G^{r - 1}) - r_1 r_3 (G^{-} - G^{r - 1})
}.
\end{equation}

Let $\Delta W_{c_i}^r = -\lambda \sum_{e = 1}^E \nabla F_{c_i} (W_{c_i}^{r, e-1} ; \xi_{c_i}^{r, e} )$, we plug it into Eq.~\eqref{eq:average_median_norm} and Eq.~\eqref{eq:limit_norm} and have
\begin{equation}\label{eq:benign_diff}
\rseq{0.72\linewidth}{!}{
\begin{aligned}
G^{+} - G^{r - 1} &= \sum_{c_i \in \mathcal{C}^{+}}-\nu_{c_i} p_{c_i} \lambda \sum_{e=1}^E \nabla F_{c_i}\left(W_{c_i}^{e-1} ; \xi_{c_i}^e\right) \\
&\stackrel{(a)}{\leq} - \lambda \sum_{c_i \in \mathcal{C}^{+}} \sum_{e=1}^E \nabla F_{c_i}\left(W_{c_i}^{e-1} ; \xi_{c_i}^e\right)
\end{aligned}
},
\end{equation}
where $\nu_{c_i} = \min (1, \nicefrac{n^r}{\|\Delta W_{c_i}^r\|_2})$ and $p_{c_i} = \nicefrac{e^{\bs{d}^{+}_i}}{\sum_j e^{\bs{d}^{+}_j}}$.
Inequality (a) is due to $\nu_{c_i} \leq 1$, $p_{c_i} \geq 0$ and $\sum_{c_i \in \mathcal{C}^{+}} p_{c_i} = 1$.
Similarly, we have
\begin{equation}\label{eq:malicious_diff}
\rseq{0.70\linewidth}{!}{
\begin{aligned}
G^{-} - G^{r - 1} &= - \lambda \sum_{c_i \in \mathcal{C}^{-}} \sum_{e=1}^E \nabla F_{c_i}\left(\tilde{W}_{c_i}^{e-1} ; \tilde{\xi}_{c_i}^e\right)
\end{aligned}
}.
\end{equation}

Finally, by plugging Eq.~\eqref{eq:benign_diff} and Eq.~\eqref{eq:malicious_diff} into Eq.~\eqref{eq:diff_proof}, Lemma~\ref{lemma:diff} is proved.

\subsection{Proof of Lemma~\ref{lemma:inner_product}}

Using Eq.~\eqref{eq:diff_lemma_eq}, we have
\begin{equation}\label{eq:lemma_inner_product_formula_1}
\rseq{0.90\linewidth}{!}{
\begin{aligned}
H_1 & \stackrel{(b)}{\leq} - r_4 \sum_{c_i \in \mathcal{C}^{+}} \sum_{e=1}^E \mathbb{E}\left[\left\langle\nabla F\left(G^{r-1}\right), \nabla F_{c_i}\left(W_{c_i}^{e-1} ; \xi_{c_i}^e\right)\right\rangle\right] + \\
& \qquad r_5 \sum_{c_i \in \mathcal{C}^{-}} \sum_{e = 1}^E \mathbb{E}\left[\left\langle\nabla F\left(G^{r-1}\right), \nabla F_{c_i}\left(\tilde{W}_{c_i}^{e-1} ; \tilde{\xi}_{c_i}^e\right)\right\rangle\right] \\
&\stackrel{(c)}{\leq} \frac{r_4}{2} \sum_{c_i \in \mathcal{C}^{+}} \sum_{e=1}^E\left(\mathbb{E}\left[\left\|\nabla F\left(G^{r-1}\right)-\nabla F_{c_i}\left(W_{c_i}^{e-1} ; \xi_{c_i}^e\right)\right\|^2\right] - H_3\right) + \\
& \qquad \frac{r_5}{2} \sum_{c_i \in \mathcal{C}^{-}} \sum_{e=1}^E\left(H_3 + \mathbb{E}\left[\left\|\nabla F_{c_i}\left(\tilde{W}_{c_i}^{e-1} ; \tilde{\xi}_{c_i}^e\right)\right\|^2\right]\right) \\
& \leq \frac{r_5 - r_4}{2} K E H_3 + \frac{r_4}{2} \sum_{c_i \in \mathcal{C}^{+}} \sum_{e=1}^E \underbrace{\mathbb{E}\left[\left\|\nabla F\left(G^{r-1}\right)-\nabla F_{c_i}\left(W_{c_i}^{e-1} ; \xi_{c_i}^e\right)\right\|^2\right]}_{H_{10}} + \\
&\qquad \frac{r_5}{2} \sum_{c_i \in \mathcal{C}^{-}} \sum_{e = 1}^E \underbrace{\mathbb{E}\left[\left\|\nabla F_{c_i}\left(\tilde{W}_{c_i}^{e-1} ; \tilde{\xi}_{c_i}^e\right)\right\|^2\right]}_{H_{11}}
\end{aligned}
},
\end{equation}
where inequality $(b)$ is obtained by Lemma~\ref{lemma:inner_product}, and inequality $(c)$ is due to $\langle\bs{x}_1, \bs{x}_2 \rangle= \nicefrac{1}{2}( \| \bs{x}_1 \|^2 + \|\bs{x}_2 \|^2 - \| \bs{x}_1 - \bs{x}_2 \|^2 )$.

In Eq.~\eqref{eq:lemma_inner_product_formula_1}, the term $H_{10}$ can be further bounded as
$$
\rseq{0.65\linewidth}{!}{
\begin{aligned}
H_{10} &\stackrel{(d)}{\leq} 2 \mathbb{E}\left[\left\|\nabla F\left(G^{r-1}\right)-\nabla F_{c_i}\left(G^{r-1}\right)\right\|^2\right] + \\
& \quad 2 \mathbb{E}\left[\left\|\nabla F_{c_i}\left(G^{r-1}\right)-\nabla F_{c_i}\left(W_{c_i}^{e-1} ; \xi_{c_i}^e\right)\right\|^2\right] \\
& \stackrel{(e)}{\leq} 2 D_i^2 + 2 L^2 H_4
\end{aligned}
},
$$
where inequality $(d)$ is due to $\|\bs{x}_1 + \bs{x}_2\|^2 \leq 2\|\bs{x}_1\|^2+2\|\bs{x}_2\|^2$, and inequality $(e)$ is due to Assumption~\ref{assump:smooth}.

Similarly, the term $H_{11}$ is bounded as
\begin{equation}\label{eq:term_11}
\rseq{0.50\linewidth}{!}{
\begin{aligned}
H_{11} \leq \zeta^2 + 2 L^2 H_5 + 4 D_i^2 + 4 H_3
\end{aligned}
}.
\end{equation}

Finally, by substituting the above equations into Eq.~\eqref{eq:lemma_inner_product_formula_1}, we can obtain Lemma~\ref{lemma:inner_product} directly.

\subsection{Proof of Lemma~\ref{lemma:diff_norm}}

According to Lemma~\ref{lemma:diff}, we have
\begin{equation}
\rseq{0.99\linewidth}{!}{
\begin{aligned}
H_2 &\leq 2 r_4^2 K E \sum_{c_i \in \mathcal{C}^{+}} \sum_{e = 1}^E \underbrace{\mathbb{E}\left[\left\|\nabla F_{c_i}\left(W_{c_i}^{e-1} ; \xi_{c_i}^e\right)\right\|^2\right]}_{H_{12}} + 2 r_5^2 K E \sum_{c_i \in \mathcal{C}} \sum_{e = 1}^E H_{11} \\
& \stackrel{(f)}{\leq} 2 r_4^2 K E \sum_{c_i \in \mathcal{C}^{+}} \sum_{e = 1}^E \left[ \zeta^2 + 2 L^2 H_4 + 4 D_i^2 + 4 H_3 \right] + \\
&\qquad 2 r_5^2 K E \sum_{c_i \in \mathcal{C}^{-}} \sum_{e = 1}^E \left[ \zeta^2 + 2 L^2 H_5 + 4 D_i^2 + 4 H_3 \right] 
\end{aligned}
},
\end{equation}
where inequality $(f)$ is due to Eq.~\eqref{eq:term_11}.
Given that the maximal number of clients within $\mathcal{C}^{+}$(or $\mathcal{C}^{-}$) is $K$, we can obtain Lemma~\ref{lemma:diff_norm}.

\subsection{Proof of Lemma~\ref{lemma:diff_norm_local}}

We have
\begin{equation}\label{eq:term_H4}
\rseq{0.99\linewidth}{!}{
\begin{aligned}
H_4 &= \mathbb{E}\left[\left\|\lambda \sum_{t = 1}^{e-1} \nabla F_{c_i}\left(W_{c_i}^{t-1} ; \xi_{c_i}^t\right)\right\|^2\right] \\
& \leq \lambda^2(e-1) \sum_{t = 1}^{e-1} \mathbb{E}\left[\left\|\nabla F_{c_i}\left(W_{c_i}^{t-1} ; \xi_{c_i}^t\right)\right\|^2\right] \\
& \stackrel{(g)}{\leq} \lambda^2 E^2 \zeta^2 + \lambda^2(e-1) \sum_{t = 1}^{e-1} \mathbb{E}\left[\left\|\nabla F_{c_i}\left(W_{c_i}^{t-1}\right)\right\|^2\right] \\
& \stackrel{(h)}{\leq} \lambda^2 E^2 \zeta^2 + 2 \lambda^2(e-1) \sum_{t=1}^{e-1} \mathbb{E}\left[\left\|\nabla F_{c_i}\left(W_{c_i}^{t-1}\right)-\nabla F_{c_i}\left(G^{r-1}\right)\right\|^2\right]+ \\
& \quad 2 \lambda^2(e-1) \sum_{t-1}^{e-1} \mathbb{E}\left[\left\|\nabla F_{c_i}\left(G^{r-1}\right)\right\|^2\right] \\
& \leq \lambda^2 E^2 \zeta^2 + 2 \lambda^2 E L^2 \sum_{t=1}^{e-1} \mathbb{E}\left[\left\| W_{c_i}^{t-1} - G^{r-1}\right\|^2\right]+ 4 \lambda^2 E^2 D_i^2 + 4 \lambda^2 E^2 H_3
\end{aligned}
},
\end{equation}
where inequality $(g)$ is due to $\mathbb{E} [\lVert \bs{x} \rVert^2] = \mathbb{E} [\lVert \bs{x} - \mathbb{E} [\bs{x}]\rVert^2] + \lVert \mathbb{E} [\bs{x} ] \rVert^2$ and $\mathbb{E} [\nabla F_{c_i}\left(W_{c_i}^{t-1} ; \xi_{c_i}^t\right)] = \mathbb{E} [\nabla F_{c_i}\left(W_{c_i}^{t-1}\right)]$, and inequality $(h)$ is due to $\mathbb{E} [\lVert \bs{x} + \bs{y} \rVert^2] \leq 2 \mathbb{E} [\lVert\bs{x}\rVert^2] + 2 \mathbb{E} [\lVert\bs{y}\rVert^2]$.
Then, summing both sides of Eq.~\eqref{eq:term_H4} from $e = 1$ to $E$ yields
\begin{equation}\label{eq:term_H4_second}
\rseq{0.80\linewidth}{!}{
\begin{aligned}
\sum_{e = 1}^E H_4 &\leq \lambda^2 E^3 \zeta^2 + 2 \lambda^2 E L^2 \sum_{e = 1}^E \sum_{t=1}^{e-1} \mathbb{E}\left[\left\|W_{c_i}^{t-1}-G^{r-1}\right\|^2\right]+ \\
&\qquad 4 \lambda^2 E^3 D_i^2+4 \lambda^2 E^3 H_3 \\
&\stackrel{(i)}{\leq} \lambda^2 E^3 \zeta^2+2 \lambda^2 E^2 L^2 \sum_{e=1}^E H_4 + 4 \lambda^2 E^3 D_i^2+4 \lambda^2 E^3 H_3
\end{aligned}
},
\end{equation}
where inequality $(i)$ is because the occurrence number of $\|W_{c_i}^{t-1}-G^{r-1}\|^2$ for each $e \in [1, E]$ in the second term is less than the number of local updating steps $E$.

\SetAlgoSkip{SkipBeforeAndAfter}
\begin{algorithm}[!t]
    \SetKwInput{KwInput}{Input}
    \SetKwInput{KwOutput}{Output}
    \SetKwInput{KwParam}{Parameter}
    \caption{G$^2$uardFL}\label{alg:guardfl}

    \KwInput{
        Number of clients $M$,
        Initial global model $G^0$,
        Number of FL rounds $R$
    }
    \KwParam{
        Client embedding training epochs $E_p$,
        Graph clustering training epochs $E_c$
    }
    \KwOutput{
        Final global model $G^R$
    }

    \SetAlgoVlined
    Initialize benign scores $\bs{s}^{0} \sim N(0, {10}^{-3})$  \;
    \For{$r \leftarrow 1$ \KwTo $R$}{   \label{alg_inner:fl_round}

        Clients $\mathcal{C}_m^r \leftarrow$ sampled from $\{c_1, \cdots, c_{M} \}$ \;

        \For{$c_i \in \mathcal{C}_m^r$}{
            $W_{c_i}^r \leftarrow$ ClientUpdate($G^{r - 1}$) \;
        }

        Constructing $\mathcal{G}^r = (\mathcal{C}, \mathbf{E}^r, \mathbf{X}^r)$ via Eq.~\eqref{eq:cos_edge} \;

        Initialize $Q$ via $\opn{HDBSCAN}(\hat{\mathbf{X}}^r)$ \;
        \For{$e \leftarrow 1$ \KwTo $E_p$}{
            Pre-training encoder $\opn{Enc}(\cdot, \cdot)$ via Eq.~\eqref{eq:loss_recon} \;
        }
        Initialize cluster center matrix $\mathbf{Q}$ via $\opn{K-Means} (\mathbf{Z})$ and Eq.~\eqref{eq:graph_encode} \;
        \For{$e \leftarrow 1$ \KwTo $E_c$}{
            Training $\opn{Enc}(\cdot, \cdot)$ and $\mathbf{Q}$ via Eq.~\eqref{eq:total_loss} \;
        }

        Computing benign cluster $\mathcal{C}^{+}$ and malicious cluster $\mathcal{C}^{-}$ via Eq.~\eqref{eq:pick_up_client} and Eq.~\eqref{eq:filter_client} \;

        Computing $\overline{W}_{c_i}^r$ to limit norms of model updates via Eq.~\eqref{eq:limit_norm} \;
        Computing benign global model and malicious global model via Eq.~\eqref{eq:aggregate_benign_model} \;
        Computing aggregated model $G^r$ via Eq.~\eqref{eq:aggregate_global} \;

        Updating benign scores via $\bs{s}^{r + 1} = \opn{tanh} \left( \bs{s}^r \right)$ \;
    }

    \KwRet{$G^R$}
    
\end{algorithm}

Similarly, we can obtain
\begin{equation}\label{eq:term_H5_second}
\rseq{0.80\linewidth}{!}{
\begin{aligned}
\sum_{e = 1}^E H_5 & \leq \lambda^2 E^3 \zeta^2+2 \lambda^2 E^2 L^2 \sum_{e=1}^E H_5 + 4 \lambda^2 E^3 D_i^2+4 \lambda^2 E^3 H_3
\end{aligned}
}.
\end{equation}

Finally, rearranging the terms in Eq.~\eqref{eq:term_H4_second} and Eq.~\eqref{eq:term_H5_second} yields Lemma~\ref{lemma:diff_norm_local}.

\begin{table}[t]
    \centering
    \caption{Measures of statistical characteristics.}\label{tab:stats_method}
    \renewcommand\arraystretch{1.5}
    \scalebox{0.70}{
    \begin{threeparttable}
        \begin{tabular}{p{0.35cm}<{\centering} p{0.95cm}<{\centering} p{3.35cm}|p{0.35cm}<{\centering} p{0.95cm}<{\centering} p{3.35cm}}
            \toprule[1.2pt]
            \textbf{\#} & \textbf{Feature} & \makecell[c]{\textbf{Description}} & \textbf{\#} & \textbf{Feature} & \makecell[c]{\textbf{Description}} \\
            \midrule[1pt]
            1 & $cos^{1}$ & Cosine distance & 2 & \emph{norm} & $L_p$ norm of vector, e.g., $L_2$ \\
            \hline
            3 & \emph{min} & Minimum vector element & 4 & \emph{max} & Maximum vector element \\
            \hline
            5 & \emph{mean} & Mean of vector elements & 6 & \emph{std} & Standard deviation of vector \\
            \hline
            7 & \emph{sum} & Sum of vector elements & 8 & \emph{median} & Median vector element \\
            \hline
            9 & $\opn{P_5}$ & 5-th percentile of elements & 10 & $\opn{P_{95}}$ & 95-th percentile of elements \\
            \bottomrule[1.2pt]
        \end{tabular}
        \begin{tablenotes}
            \item[1] : The measure is applied to two vectors.
        \end{tablenotes}
    \end{threeparttable}
    }
    \vspace{-10pt}
\end{table}

\section{Details of Experimental Setup}

\subsection{Algorithm}\label{subsec:algorithm}

The measures of statistical characteristics adopted to construct clients' features are listed in Table~\ref{tab:stats_method}.

In this subsection, we present a comprehensive description of the G$^2$uardFL process, detailed in Algorithm~\ref{alg:guardfl}.
In Lines 3 - 4, clients receive the global model $G^{r - 1}$ from the previous round and utilize their local data to train their individual models, represented as $W_{c_i}^r$.
The algorithm employs two standard techniques, HDBSCAN, and K-means, to initialize the number of cluster centroids $Q$ (Line 7) and cluster centers $\mathbf{Q}$ (Line 10), respectively.
Note that HDBSCAN takes the client feature matrix $\hat{\mathbf{E}}^r$ as input, whereas K-means operates on the latent representation matrix $\mathbf{Z}$.

\subsection{Dataset}\label{subsec:detail_dataset}

In our evaluations, we employ four distinct datasets: MNIST, CIFAR-10, Sentiment-140, and Reddit.
Following Bagdasaryan \emph{et al}.~\cite{bagdasaryan2020how}, we consider only the first 100 and 40 words of sentences in the Sentiment-140 and Reddit datasets, respectively.
If a sentence falls below this threshold, we pad it with the \emph{\textless pad\textgreater} symbol to reach the required length.
The process of creating backdoored samples aligns with Wang \emph{et al}.\cite{wang2020attack}.

For MNIST, we apply a slightly modified LeNet-5 architecture~\cite{wang2020attack}.
For CIFAR-10, we employ a 9-layer VGG-style network architecture, with the removal of all BatchNorm layers.
This adjustment is necessary to mitigate the potential adverse effects of poorly managed BatchNorm layers on primary task performance, particularly with small batch sizes and non-iid data~\cite{sattler2019robust}.
For Sentiment-140, our model consists of an embedding layer ($V \times 50$) and a 2-layer LSTM with 32 hidden dimensions, followed by a linear layer ($64 \times 2$) and a Sigmoid activation.
In this dataset, the vocabulary size $V$ is $122971$.
For Reddit, the model incorporates an embedding layer ($V \times 50$) and a 2-layer LSTM with 64 hidden dimensions, followed by a linear layer ($128 \times V$) and a softmax activation, with $V$ being $49999$.

\begin{figure}[t]
    \centering
    \includegraphics[width=0.99\linewidth]{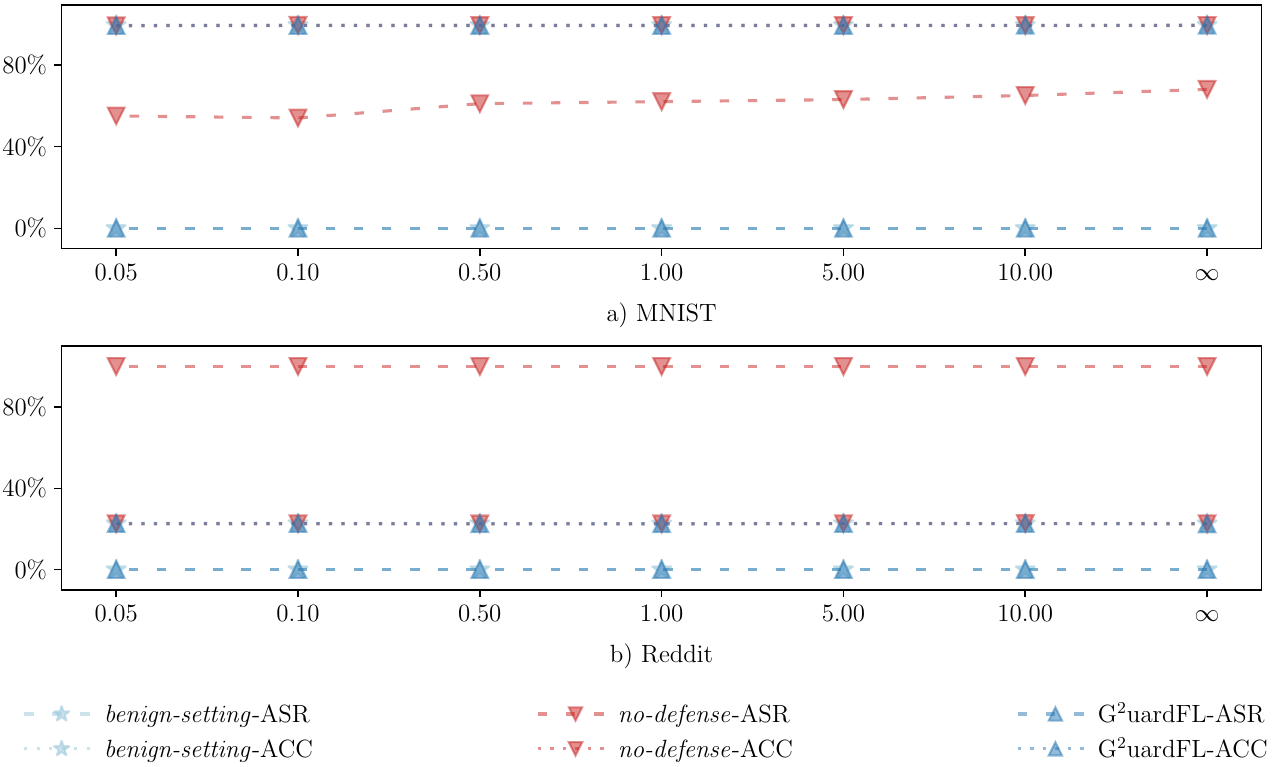}
    \vspace{-10pt}
    \caption{The Impact of data heterogeneity on G$^2$uardFL in MNIST and Reddit.}\label{fig:mnist_and_reddit_data}
    \vspace{-5pt}
\end{figure}

\begin{figure}[t]
    \centering
    \includegraphics[width=0.99\linewidth]{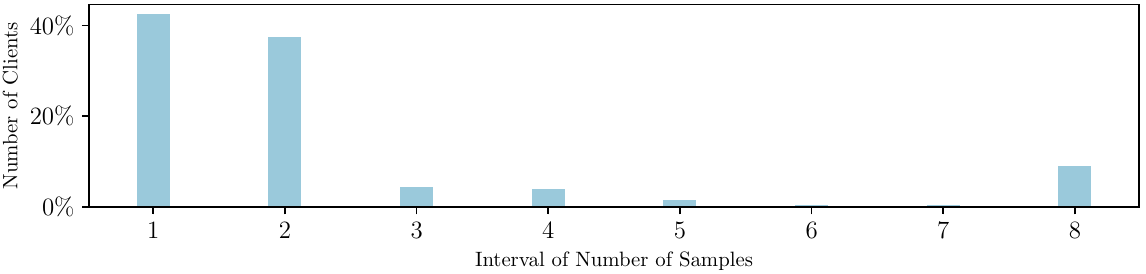}
    \vspace{-10pt}
    \caption{Distribution of data volume on different clients.}\label{fig:vis_sentiment}
    \vspace{-10pt}
\end{figure}

\subsection{Baselines}\label{subsec:detail_baseline}

We benchmark G$^2$uardFL against nine state-of-the-art defenses, and each defense with specific configurations is outlined as follows:
1) \textbf{FLAME}~\cite{nguyen2022flame}.
Noise level factor $\lambda$ to 0.001 for all tasks.
2) \textbf{DeepSight}~\cite{rieger2022deep}.
The threshold for excluding suspicious models from clusters is set at 0.33, and the number of fake data used to calculate DDIFs is set to 2,000.
3) \textbf{FoolsGold}~\cite{fung2020limitations}.
The confidence parameter $\kappa$ is set to 1, and it does not employ the historical gradient or the significant feature filter.
4) \textbf{Krum}~\cite{blanchard2017machine}.
The local model with the highest score is selected as the aggregated global model.
5) \textbf{Multi-Krum}~\cite{blanchard2017machine}.
The number of clients participating per round is set to 10, the number of permissible malicious clients $f$ is 2, and the hyper-parameter $m$ is set to $n - f$.
6) \textbf{RLR}~\cite{ozdayi2021defending}.
We set the learning rate threshold $\theta = 3$ and the server's learning rate $\eta = 1$.
7) \textbf{FLTrust}~\cite{cao2021fltrust}.
We set the number of samples within the server to 1\% of the validation dataset in all evaluations.
8) \textbf{NDC}~\cite{sun2019can}.
The threshold of norm difference between the global model and model updates is set to $2$ for all tasks.
9) \textbf{RFA}~\cite{pillutla2019robust}.
We set the smoothing factor $v = 10^{-5}$, the maximum number of iterations $T = 500$ and the fault tolerance threshold $\epsilon = 10^{-1}$.
10) \textbf{Weak DP}~\cite{dwork2014algorithmic}.
The noise level $\sigma$ is 0.025.

\subsection{Backdoor Attacks}\label{subsec:detail_backdoor_attack}

We investigate the robustness of G$^2$uardFL against five distinct backdoor attacks, each configured as follows:
1) \textbf{Black-box Attack}~\cite{wang2020attack}.
Malicious clients train on backdoored samples $\mathcal{D}^{\mathcal{A}}$ using the same hyper-parameters (e.g., learning rate) as honest clients for all tasks.
2) \textbf{PGD without replacement}~\cite{wang2020attack}.
Malicious clients train on backdoored samples $\mathcal{D}^{\mathcal{A}}$, projecting poisoned weights onto an $L_2$ ball of radius $\epsilon = \{0.2, 1.0, 2.0, 2.0\}$ for MNIST, CIFAR-10, Sentiment-140, and Reddit, respectively.
3) \textbf{PGD with replacement}~\cite{wang2020attack}.
Configuration is aligned with PGD without replacement.
4) \textbf{Constrain-and-scale}~\cite{bagdasaryan2020how}.
The factor $\alpha$, which balances the effectiveness and stealthiness, is set to $0.5$.
5) \textbf{DBA}~\cite{xie2020dba}.
The trigger factors are set to $\phi = \{4, 2, 0\}$ for MNIST and $\phi = \{6, 3, 0\}$ for CIFAR-10.
6) \textbf{3DFed}~\cite{li20233dfed}. $\kappa$, a scaling factor in the indicator, is set as a constant $10^{5}$.

\begin{table}[t]
    \centering
    \caption{Results of ablation study (\%).}\label{tab:ablation_study}
    \renewcommand\arraystretch{1.5}
    \scalebox{0.64}{
    \begin{threeparttable}
        \begin{tabular}{p{1.2cm} p{1.0cm}<{\centering} p{1.0cm}<{\centering} p{1.0cm}<{\centering} p{1.0cm}<{\centering} p{1.0cm}<{\centering} p{1.0cm}<{\centering} p{1.0cm}<{\centering} p{1.0cm}<{\centering}}
            \toprule[1.2pt]
            \multirow{3}{*}{\parbox{1.2cm}{\centering Defense}} & \multicolumn{4}{c}{MNIST} & \multicolumn{4}{c}{CIFAR-10} \\
            ~ & \multicolumn{2}{c}{$\alpha = \infty$} & \multicolumn{2}{c}{$\alpha = 0.05$} & \multicolumn{2}{c}{$\alpha = \infty$} & \multicolumn{2}{c}{$\alpha = 0.05$} \\
            ~ & ASR & ACC & ASR & ACC & ASR & ACC & ASR & ACC \\
            \midrule[1pt]
            G$^2$uardFL & 00.00 & 99.33 & 00.00 & 99.19 & 01.67 & 84.84 & 01.11 & 80.19 \\
            G$^2$uardFL${}^\prime$ & 00.00 & 99.32 & 05.00 & 99.21 & 02.22 & 84.38 & 79.44 & 79.93 \\
            \bottomrule[1.2pt]
            \multirow{3}{*}{\parbox{1.2cm}{\centering Defense}} & \multicolumn{4}{c}{Sentiment-140} & \multicolumn{4}{c}{Reddit} \\
            ~ & \multicolumn{2}{c}{$\alpha = \infty$} & \multicolumn{2}{c}{$\alpha = 0.05$} & \multicolumn{2}{c}{$\alpha = \infty$} & \multicolumn{2}{c}{$\alpha = 0.05$} \\
            ~ & ASR & ACC & ASR & ACC & ASR & ACC & ASR & ACC \\
            \midrule[1pt]
            G$^2$uardFL & 07.92 & 74.12 & 43.56 & 73.00 & 00.00 & 22.48 & 00.00 & 22.68 \\
            G$^2$uardFL${}^\prime$ & 31.28 & 72.73 & 100.00 & 62.27 & 00.00 & 22.84 & 83.51 & 22.83 \\
            \bottomrule[1.2pt]
        \end{tabular}
    \end{threeparttable}
    }
    \vspace{-5pt}
\end{table}

\section{Experimental Results}

\subsection{Data Heterogeneity}\label{subsec:appendix_data_heterogeneity}

We present the experimental results of data heterogeneity for the MNIST and Reddit datasets in Fig.~\ref{fig:mnist_and_reddit_data}.
We have observed a significant decrease in G$^2$uardFL's performance on the Sentiment-140 dataset when data heterogeneity is set to $0.05$.
To further investigate this, we analyze the distribution of data volume across different clients on Sentiment-140 with $\alpha = 0.05$ across eight intervals: $[1, 2)$, $[2, 500)$, $[500, 1000)$, $[1000, 1500)$, $[1500, 2000)$, $[2000, 2500)$, $[2500, 3000)$, and $[3000, \infty)$, as depicted in Fig.~\ref{fig:vis_sentiment}.
We observe that over 40\% of clients have only one sample.
Consequently, model weights of benign clients are not sufficiently trained with normal data, leading to a degradation in primary task performance.
Furthermore, if poisoned weights are mistakenly accepted by the server, purging the global model of embedded backdoors by training local models with client local data becomes challenging.

\subsection{Analysis of Adaptive Poison Eliminating}\label{subsec:ada_impact}

\begin{table}[t]
    \centering
    \caption{Results of convergence analysis (\%).}\label{tab:convergence_analysis}
    \renewcommand\arraystretch{1.5}
    \scalebox{0.64}{
    \begin{threeparttable}
        \begin{tabular}{p{1.4cm} p{2.4cm}<{\centering} p{2.4cm}<{\centering} p{2.4cm}<{\centering} p{2.4cm}<{\centering}}
            \toprule[1.2pt]
            \multirow{2}{*}{\parbox{1.4cm}{\centering Defense}} & \multicolumn{2}{c}{MNIST} & \multicolumn{2}{c}{CIFAR-10} \\
            ~ & $\alpha = \infty$ & $\alpha = 0.05$ & $\alpha = \infty$ & $\alpha = 0.05$ \\
            \midrule[1pt]
            no-defense & 99.19 & 98.74 & 81.79 & 33.23 \\
            G$^2$uardFL & 98.90 & 97.37 & 77.67 & 32.31 \\
            \bottomrule[1.2pt]
            \multirow{2}{*}{\parbox{1.4cm}{\centering Defense}} & \multicolumn{2}{c}{MNIST} & \multicolumn{2}{c}{CIFAR-10} \\
            ~ & $\alpha = \infty$ & $\alpha = 0.05$ & $\alpha = \infty$ & $\alpha = 0.05$ \\
            \midrule[1pt]
            no-defense & 69.87 & 57.24 & 16.30 & 16.30 \\
            G$^2$uardFL & 67.04 & 51.77 & 16.30 & 16.30 \\
            \bottomrule[1.2pt]
        \end{tabular}
    \end{threeparttable}
    }
    \vspace{-5pt}
\end{table}

\begin{figure}[!t]
    \centering
    \includegraphics[width=0.99\linewidth]{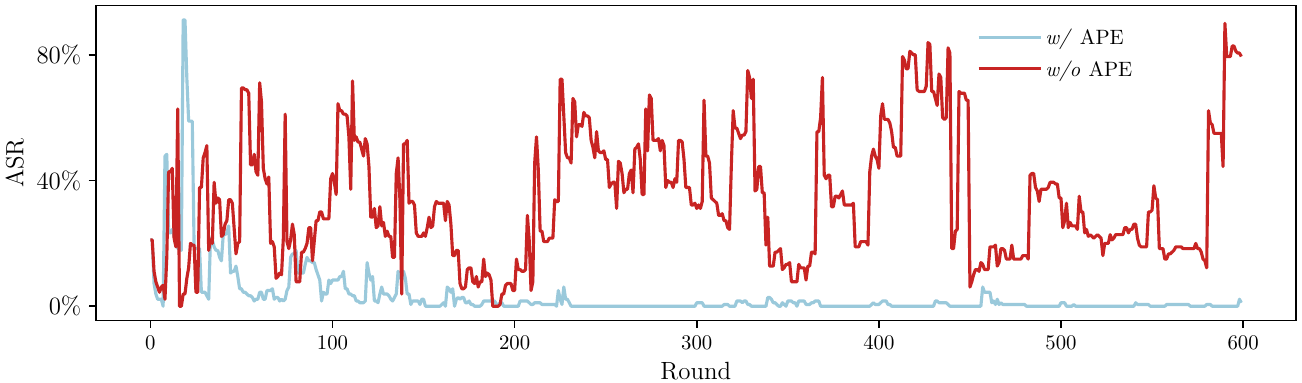}
    \vspace{-10pt}
    \caption{The ASR fluctuations on CIFAR-10.}\label{fig:vis_eliminating}
    \vspace{-10pt}
\end{figure}

In this subsection, we investigate the impact of the adaptive poison eliminating module, referred to as APE, through experiments conducted on four datasets with varying degrees of data heterogeneity, specifically $\alpha = \infty$ and $\alpha = 0.05$.
We denote G$^2$uardFL without this module as G$^2$uardFL${}^\prime$.
The results, presented in Table~\ref{tab:ablation_study}, yield several observations:

1) APE proves to be effective in swiftly eliminating previously embedded backdoors, especially in scenarios with high data heterogeneity.
The substantial disparity between model weights from benign clients increases the likelihood of accepting poisoned weights.

2) The presence of this module heightens the difficulty of launching successful backdoor attacks by augmenting the divergence between poisoned models and the aggregated global model.
Fig.~\ref{fig:vis_eliminating} illustrates the fluctuations in ASR when launching backdoor attacks on CIFAR-10 at $\alpha = 0.05$.
The ASR volatility of G$^2$uardFL with this module is significantly lower than that without it.
This can be attributed to the enlargement of the divergence between malicious and benign clients, which raises the challenge of poisoning the global model during each round, ultimately resulting in a lower ASR.

To address the concern that APE prevents the convergence of FL systems, we also conduct evaluations by training the global model from randomly initialized model weights without any attacks.
The primary task performance is listed in Table~\ref{tab:convergence_analysis}.
We observe that G$^2$uardFL only leads to a slight degradation in ACC, which remains acceptable.

\end{document}